\documentclass[twocolumn,aps,prc,superscriptaddress,showpacs,floatfix,longbibliography,nofootinbib]{revtex4-1}
\usepackage{url}
\usepackage{cancel}
\usepackage[colorlinks,linkcolor=blue,citecolor=blue,filecolor=black,urlcolor=blue]{hyperref}
\usepackage{epsfig,graphics}
\usepackage{graphicx}
\usepackage{dcolumn}
\usepackage{bm}
\usepackage[usenames]{color}
\usepackage{amssymb}
\usepackage{amsmath}
\usepackage{multirow}
\usepackage{float}
\usepackage{harpoon}
\usepackage{MnSymbol}
\usepackage{appendix}
\usepackage{color}
\usepackage{hyperref}
\usepackage{cleveref}

\begin{document}

\title{Transport properties of asymmetric nuclear matter in the spinodal region}
\author{Lei-Ming Hua}
\affiliation{Shanghai Institute of Applied Physics, Chinese Academy of Sciences, Shanghai 201800, China}
\affiliation{University of Chinese Academy of Sciences, Beijing 100049, China}
\author{Jun Xu}\email[Correspond to\ ]{junxu@tongji.edu.cn}
\affiliation{School of Physics Science and Engineering, Tongji University, Shanghai 200092, China}
\affiliation{Shanghai Institute of Applied Physics, Chinese Academy of Sciences, Shanghai 201800, China}
\begin{abstract}
We have studied the shear and bulk viscosities of asymmetric nuclear matter in the mechanical and chemical instability region based on IBUU transport simulations in a box system. The Green-Kubo method is used to calculate these viscosities with a prepared dynamically equilibrated nuclear system with hot clusters. While the behavior of the shear viscosity is largely affected by energy-dependent nucleon-nucleon cross sections, the bulk viscosity increases significantly in the presence of nuclear clusters compared to that in uniform nuclear matter. Increasing isospin asymmetry generally increases both viscosities, while their behaviors are qualitatively modified once the isospin asymmetry is large enough to affect significantly the spinodal region. Our calculation shows that the bulk viscosity is more sensitive to the nuclear clustering than the shear viscosity, and is thus a robust quantity related to the phase diagram of asymmetric nuclear matter.
\end{abstract}
\maketitle

\section{introduction}

Transport properties of nuclear matter are dominated by the shear viscosity, representing the proportional coefficient characterizing the relation between the flux gradient and the shear force per unit area, and the bulk viscosity, a dissipation coefficient due to the compressibility of the fluid~\cite{LANDAU19591}. Knowledge on transport properties of nuclear matter~\cite{Danielewicz:1984kt,Shi:2003np} may help us to understand the dynamics in heavy-ion collisions and mergers of neutron stars. The viscosity is often scaled by the entropy density, called the specific viscosity. It is a general phenomenon that both the specific shear and bulk viscosity may have non-monotonic behaviors in the vicinity of a phase transition. For instance, the specific shear viscosity has a minimum around the temperature of a hadron-quark phase transition~\cite{Csernai:2006zz,Lacey:2006bc} or a nuclear liquid-gas phase transition~\cite{Chen:2007xe,Pal:2010sj,Xu:2013nwa,Xu:2015lna,PhysRevC.105.064613}, while the specific bulk viscosity has a peak around the temperature of a phase transition~\cite{Karsch:2007jc,Kharzeev:2007wb,PhysRevC.74.014901}. In this sense, the behaviors of the transport properties are closely related to the phase diagram of strong-interacting matter.

In our previous work~\cite{Hua:2023amo}, we have studied the shear viscosity of isospin symmetric nuclear matter near its liquid-gas phase transition through the Green-Kubo method~\cite{Kubo:1957mj,Kubo_1966}, the most rigourous method among others~\cite{Plumari:2012ep}. In the present study, we further investigate the bulk viscosity and extend the approach to isospin asymmetric nuclear matter. As is known, besides the mechanical instability, chemical instability may exist in isospin asymmetric nuclear matter, leading to a richer phase structure of nuclear matter. The purpose of the present study is to understand the relation between the phase structure of isospin asymmetric nuclear matter and its transport properties, i.e., the behaviors of the shear and bulk viscosities.

\section{theoretical framework}

The present study is carried out by transport simulations in a cubic box of the size $20\times 20\times 20$ fm$^3$ with the periodic boundary condition based an isospin-dependent Boltzmann-Uehling-Uhlenbeck (IBUU) transport model,
which solves the isospin-dependent BUU equation expressed as
\begin{equation}
  \label{eq:boltz}
  \frac{\partial f_\tau(\vec{r},\vec{p})}{\partial t}
  +\frac{\vec{p}}{\sqrt{m^2+\vec{p}^2}}\cdot
  \frac{\partial f_\tau(\vec{r},\vec{p})}{\partial \vec{r}}-\frac{\partial U_\tau}{\partial \vec{r}}\cdot\frac{\partial f_\tau(\vec{r},\vec{p})}{\partial \vec{p}}
  =I_c.
\end{equation}
The above equation describes how the phase-space distribution function $f_\tau(\vec{r},\vec{p})$ for nucleons with bare mass $m$ and isospin index $\tau$ ($\tau=n,p$) evolves with time under the mean-field potential $U_\tau$ and due to collisions $I_c$.
We note both the mean-field evolution~\cite{Colonna:2021xuh} and nucleon-nucleon (NN) collisions~\cite{Zhang:2017esm} in IBUU are calibrated by the Transport Model Evaluation Project.

The mean-field evolution is based on the lattice Hamiltonian framework~\cite{Lenk:1989zz}, which maintains reasonable dynamics of nuclear clustering in the spinodal region and guarantees energy conservation. The mean-field potential in nuclear matter of density $\rho$ and isospin asymmetry $\delta$ applied in the present study is written as
\begin{eqnarray}\label{u}
U_{n,p}(\rho,\delta)=\alpha \left(\frac{\rho}{\rho_{0}}\right)+\beta\left(\frac{\rho}{\rho_{0}}\right)^{\gamma} \pm 2 E_{sym}^{pot}  \left(\frac{\rho}{\rho_{0}}\right)^{\gamma_{sym}} \delta, 
\end{eqnarray}
with the `+' (`$-$') sign for neutrons (protons). In the isoscalar part of the potential, the coefficients $\alpha = -0.218$ GeV, $\beta = 0.164$ GeV, and $\gamma = 4/3$ are fitted by the saturation density $\rho_0=0.16$ fm$^{-3}$, the binding energy $E_0=-16$ MeV at $\rho_0$, and the incompressiblity $K_0=237$ MeV. For the isovector potential, we set $E_{sym}^{pot}=18$ MeV and $\gamma_{sym}=0.656$ which lead to a nuclear symmetry energy of a value 30.3 MeV and a slope parameter $L=60$ MeV at the saturation density. We have also considered the isoscalar and isovector density gradient terms with empirical coefficients as in the MSL0 force~\cite{Chen:2010qx}, but they have minor effects on the results.

\begin{figure}[!h]
\includegraphics[width=1.0\linewidth]{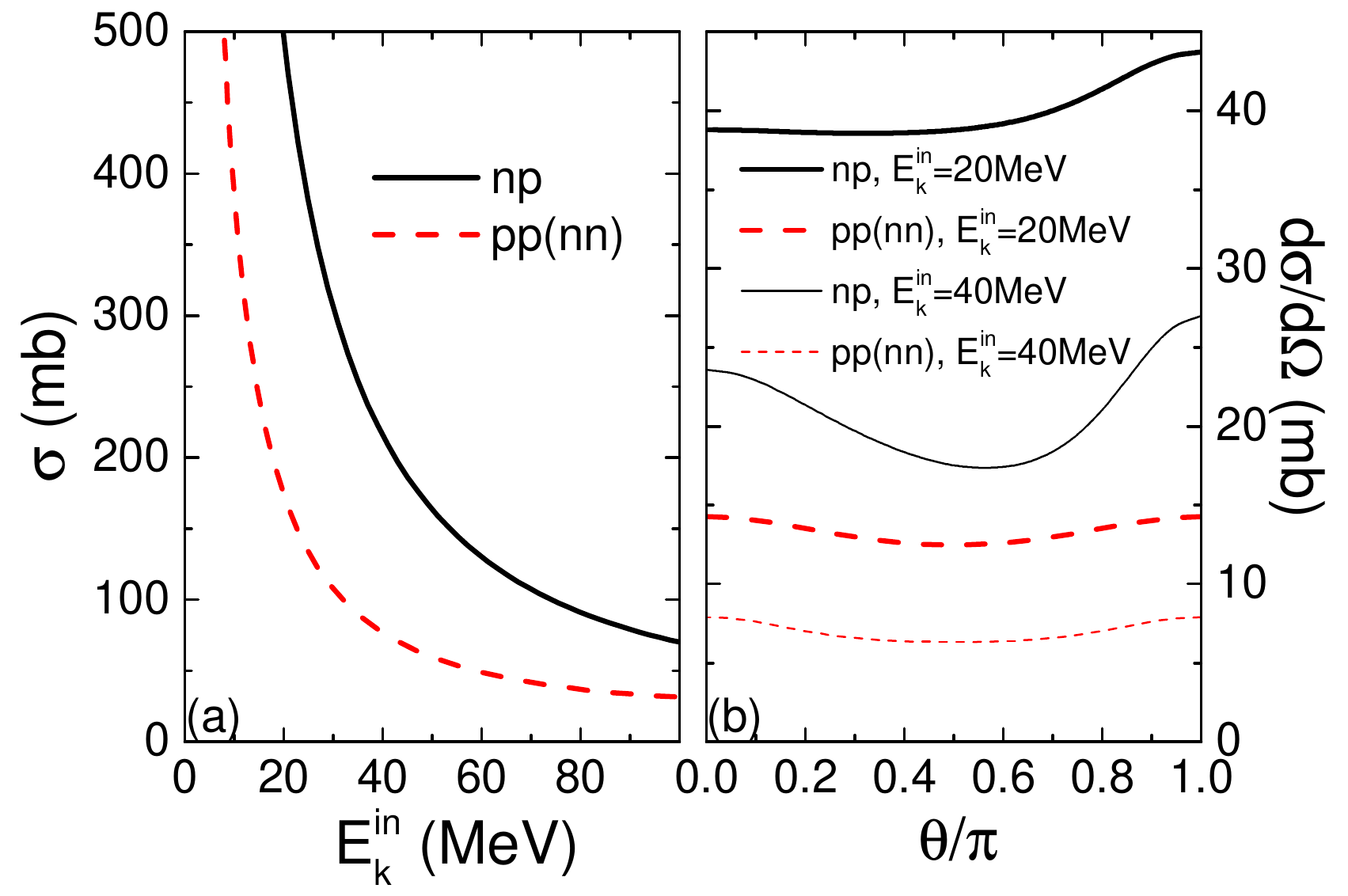}
\caption{\label{cs} Left: Total neutron-proton (np) and proton-proton (pp) cross sections as functions of the kinetic energy $E_{k}^{in}$ of incident nucleon; Right: Polar angular dependence of differential cross sections for neutron-proton (np) and proton-proton (pp) collisions at $E_{k}^{in}=20$ and 40 MeV. }
\end{figure}

The NN collisions are carried out through the modified Bertsch's prescription~\cite{Bertsch:1988ik}, basically a geometric method but with improvements made by using the time step in the center-of-mass frame of the collision pair and removing repeated collisions (see Ref.~\cite{Zhang:2017esm} for more details). Unlike our previous study~\cite{Hua:2023amo}, we use more realistic energy- and isospin-dependent differential NN collision cross sections in the present study. The energy dependence of the total cross sections and the polar angular dependence of the differential cross sections at typical incident energies are displayed in Fig.~\ref{cs}. In the present transport simulation, we adopt these free-space cross sections taken from Refs.~\cite{PhysRevC.15.1002,PhysRevC.96.044618}, since the energy- and angular-dependence of the differential in-medium cross sections are still uncertain. The very large total cross sections at low energies are automatically restricted in the finite-size box system with the periodic boundary condition, since the largest distance as well as the cross section between two nucleons has an upper limit. For the isospin-dependent Pauli blockings of NN collisions, we approximate the local phase-space distribution by a Fermi-Dirac distribution, i.e.,
\begin{equation}
f_{\tau}(\vec{r}, \vec{p})=\frac{1}{\exp \left(\frac{\sqrt{p^{2}+m^{2}}-m+U_{\tau}-\mu_{\tau}}{T}\right)+1},
\end{equation}
where the local temperature $T$ and the neutron and proton chemical potential $\mu_\tau$ can be inversely calculated from the local neutron and proton densities $\rho_\tau$ and the kinetic energy density $\epsilon_k$ from transport simulations through the relations
\begin{eqnarray}
\rho_{\tau} &=& 2\int f_{\tau}(\vec{r}, \vec{p}) \frac{d^{3} p}{(2 \pi)^{3}}, \\
\epsilon_{k} &=& \sum_{\tau} 2\int \left(\sqrt{p^{2}+m^{2}}-m\right) f_{\tau}(\vec{r}, \vec{p}) \frac{d^{3} p}{(2 \pi)^{3}}.
\end{eqnarray}

Both the shear viscosity $\eta$ and the bulk viscosity $\zeta$ can be expressed by the Green-Kubo formulas based on the fluctuation-dissipation theorem~\cite{Kubo:1957mj,Kubo_1966,Kapusta:2008vb}
\begin{eqnarray}
\eta&=&\frac{1}{T} \int d^{3} r \int_{t_0}^{\infty} d t\left\langle\pi^{x y}(\vec{0}, t_0) \pi^{x y}(\vec{r}, t)\right\rangle_{\text {equil }}, \label{eta1}\\
\zeta&=&\frac{1}{T} \int d^{3} r \int_{t_0}^{\infty} d t\left\langle \Delta \pi(\vec{0}, t_0) \Delta \pi(\vec{r}, t)\right\rangle_{\text {equil }}. \label{zeta}
\end{eqnarray}
In the above, $T$ is the temperature of the system, and $t-t_0$ is the post-equilibration time with $t_0$ being the time when the system has reached dynamic equilibrium. $\pi^{x y}$ is the shear component of the energy-momentum tensor, which can be expressed as
\begin{eqnarray}
\pi^{x y}=\frac{1}{V_c} \sum_{i} \frac{p^{x}_i p^{y}_i}{E_i}, \label{eta2}
\end{eqnarray}
where $V_c$ is the volume of the cell, and $p^{x}_i$, $p^{y}_i$, and $E_i$ are, respectively, the momentum in the $x$ and $y$ direction and the energy of the $i$th nucleon in the local cell obtained from transport simulations. $\Delta \pi=\pi-\pi_{eq}$ represents the fluctuation of the bulk component of the energy-momentum tensor
\begin{eqnarray}
\pi = \frac{1}{3} (\pi^{x x} + \pi^{y y} + \pi^{z z}) \label{pi}
\end{eqnarray}
with respect to its value $\pi_{eq}$ in the equilibrium state. In our previous work~\cite{Hua:2023amo}, we have proved that the Green-Kubo formula for the shear viscosity is valid for non-uniform nuclear matter, and Eqs.~(\ref{eta1}) and (\ref{eta2}) can be evaluated in the whole box system. Similar procedure applies for the calculation of the bulk viscosity. Therefore, the shear and bulk viscosity can be calculated from
\begin{eqnarray}
\eta&=&\frac{V}{T} \int_{t_0}^{\infty} d t\left\langle\Pi^{x y}(t_0) \Pi^{x y}(t)\right\rangle_{\text {equil }}, \label{eta0}\\
\zeta&=&\frac{V}{T} \int_{t_0}^{\infty} d t\left\langle \Delta \Pi(t_0) \Delta \Pi(t)\right\rangle_{\text {equil }}, \label{zeta0}
\end{eqnarray}
where $V$ is the box volume, and $\Pi^{x y}$ and $\Delta \Pi$ are calculated similarly as Eqs.~(\ref{eta2}) and (\ref{pi}) except that the summation is over the whole box system.

\section{results and discussions}

We will show the mechanical instability, the chemical instability, and the liquid-gas phase transition regions of isospin asymmetric nuclear matter in Sec.~\ref{a}, and present how nuclear clusters are generated in these spinodal regions in Sec.~\ref{b}. Sec.~\ref{c} discusses how the Green-Kubo method is used by fitting time evolution of the energy-momentum tensor, and gives extensive results of the shear and bulk viscosities.

\subsection{thermodynamics of isospin asymmetric nuclear matter}
\label{a}

\begin{figure}[!h]
\includegraphics[width=0.7\linewidth]{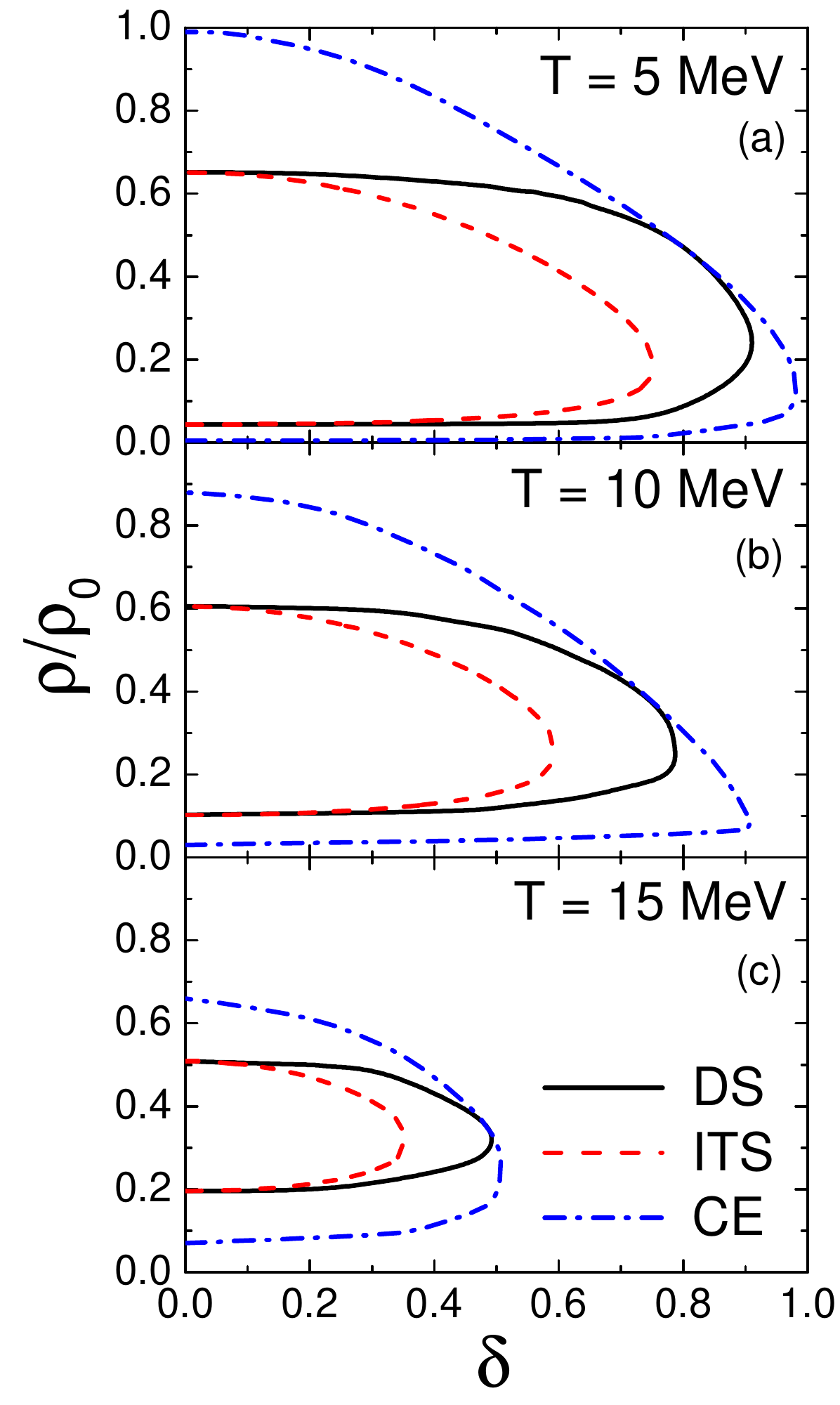}
\caption{\label{pd} Boundaries of isothermal spinodal (ITS), diffusive spinodal (DS), and liquid-gas phase coexistence (CE) in the $(\rho,\delta)$  plane at the temperatures $T=5$ (a), 10 (b), and 15 (c) MeV.}
\end{figure}

Using the mean-field potential as Eq.~(\ref{u}), we display boundaries of the mechanical instability (isothermal spinodal) region, the chemical instability (diffusive spinodal) region, and the liquid-gas phase coexistence region in the $(\rho,\delta)$ plane at different temperatures $T$ in Fig.~\ref{pd}. In the mechanical instability region with $(\partial P/\partial \rho)_{T,\delta}<0$, increasing (reducing) the local density $\rho$ reduces (increases) the local pressure $P$, so more particles will flow into (away from) the local area, further reducing (increasing) the local pressure, thus any small density fluctuations may grow and make the nuclear matter unstable. In the chemical instability region with $(\partial \mu_n/\partial \delta)_{P,T} <0$ or $(\partial \mu_p/\partial \delta)_{P,T} >0$, increasing the local isospin asymmetry $\delta$ reduces the neutron chemical potential or increases the proton chemical potential, so more neutrons will flow into the local area or more protons will flow away from the local area, further increasing the local isospin asymmetry, thus any small fluctuations of the isospin asymmetry may grow and make the nuclear matter unstable. The coexistence line describes the region where the liquid phase represented by clusters and the gas phase represented by free nucleons can coexist (see the Maxwell construction for the liquid-gas mixed phase in, e.g., Refs.~\cite{Muller:1995ji,Xu:2007eq}), and it contains both mechanical and chemical instability regions. The spinodal region shrinks with increasing temperature or increasing isospin asymmetry. While this is a thermodynamic calculation of nuclear matter, we will see the corresponding situation from IBUU simulation in a box system.

\subsection{preparation of dynamically equilibrated systems with nuclear clusters}
\label{b}

\begin{figure*}
\centering
\includegraphics[width=0.15\linewidth]{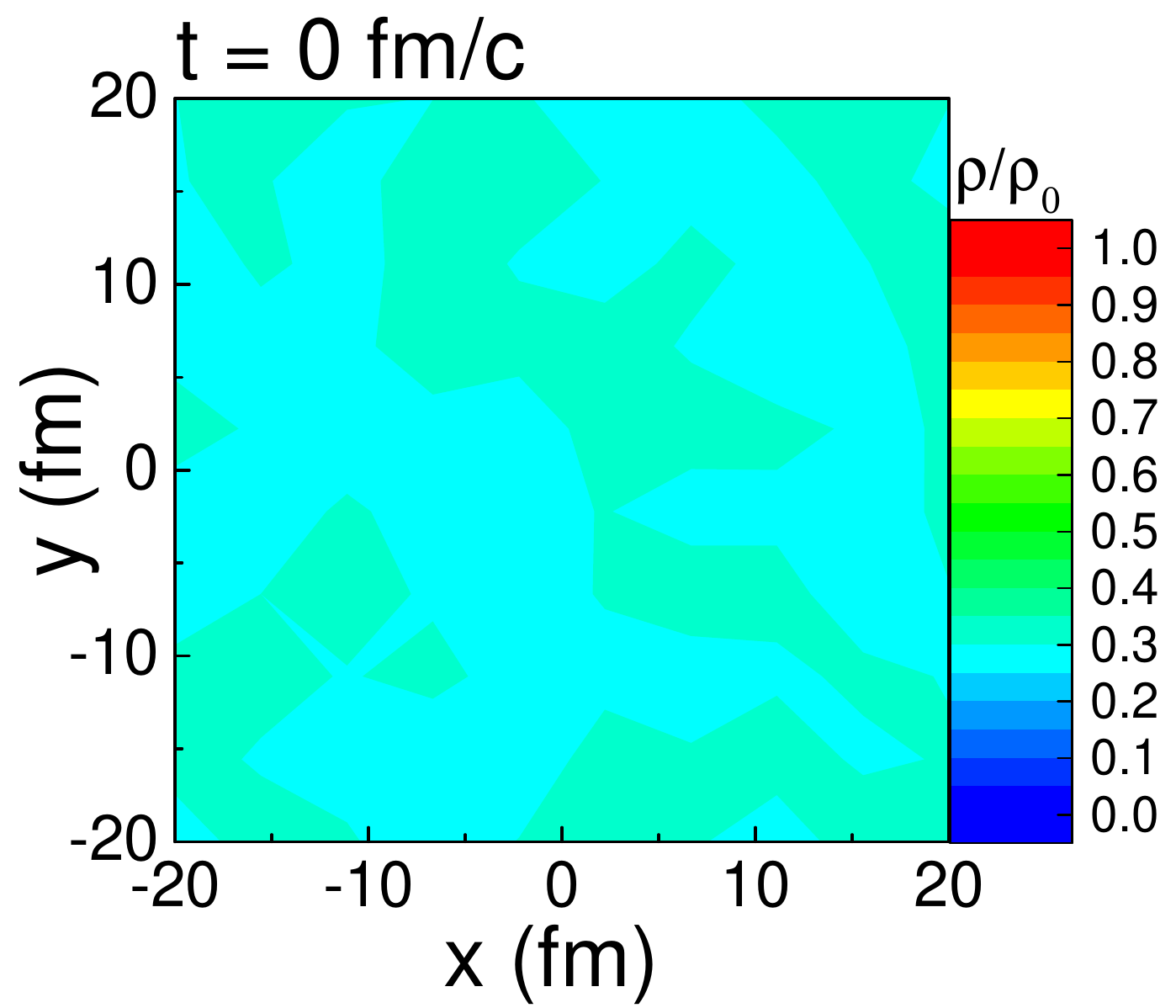}
\includegraphics[width=0.15\linewidth]{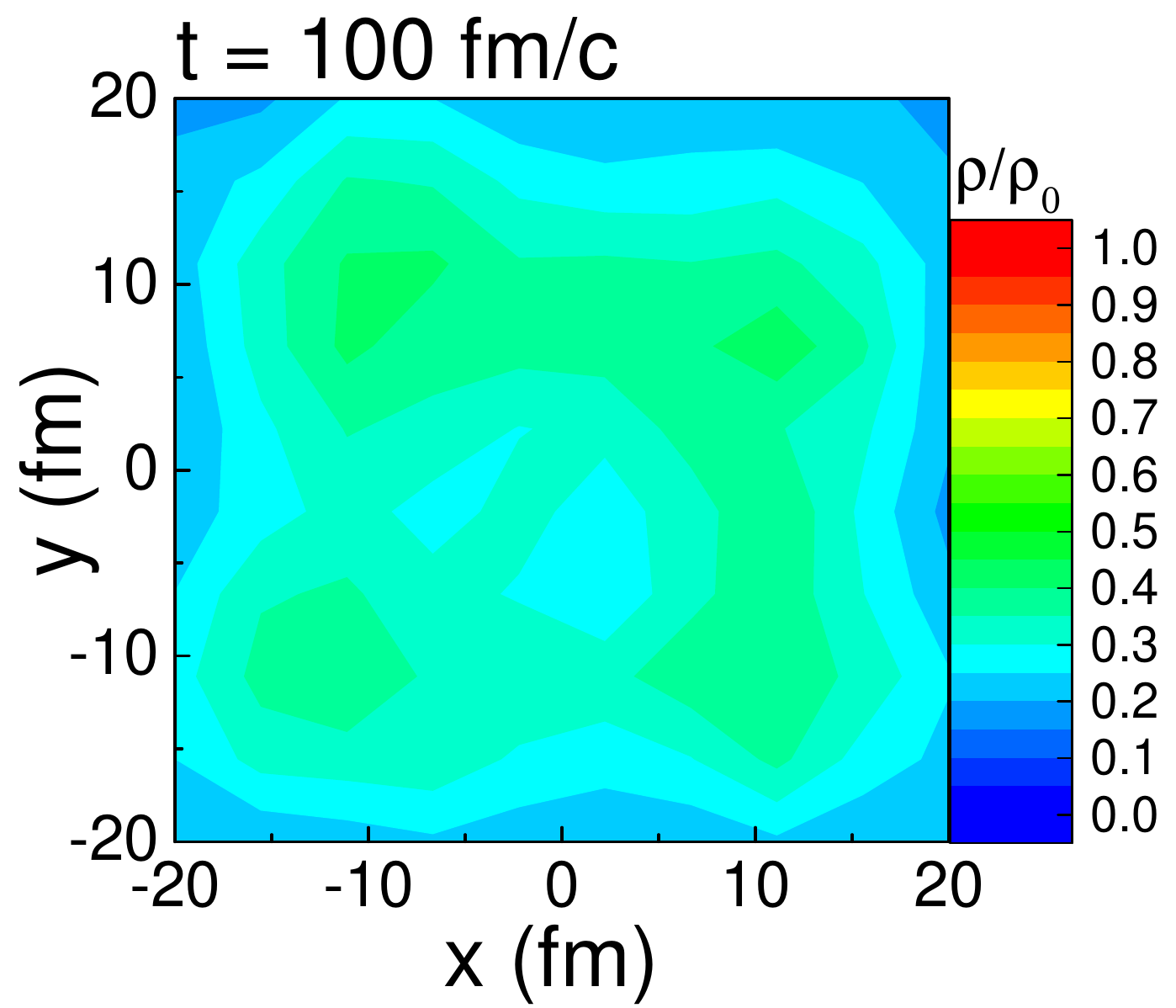}
\includegraphics[width=0.15\linewidth]{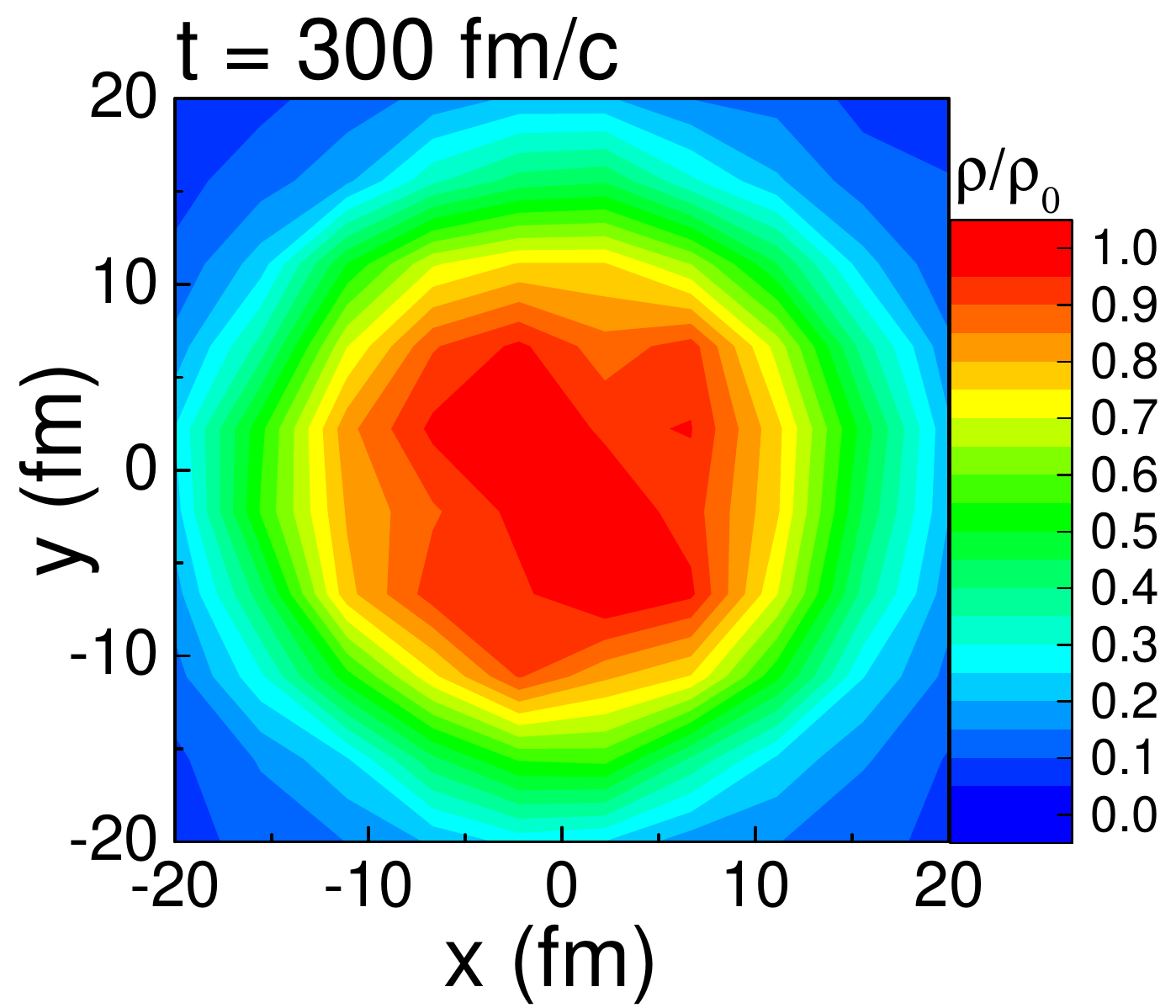}
\includegraphics[width=0.15\linewidth]{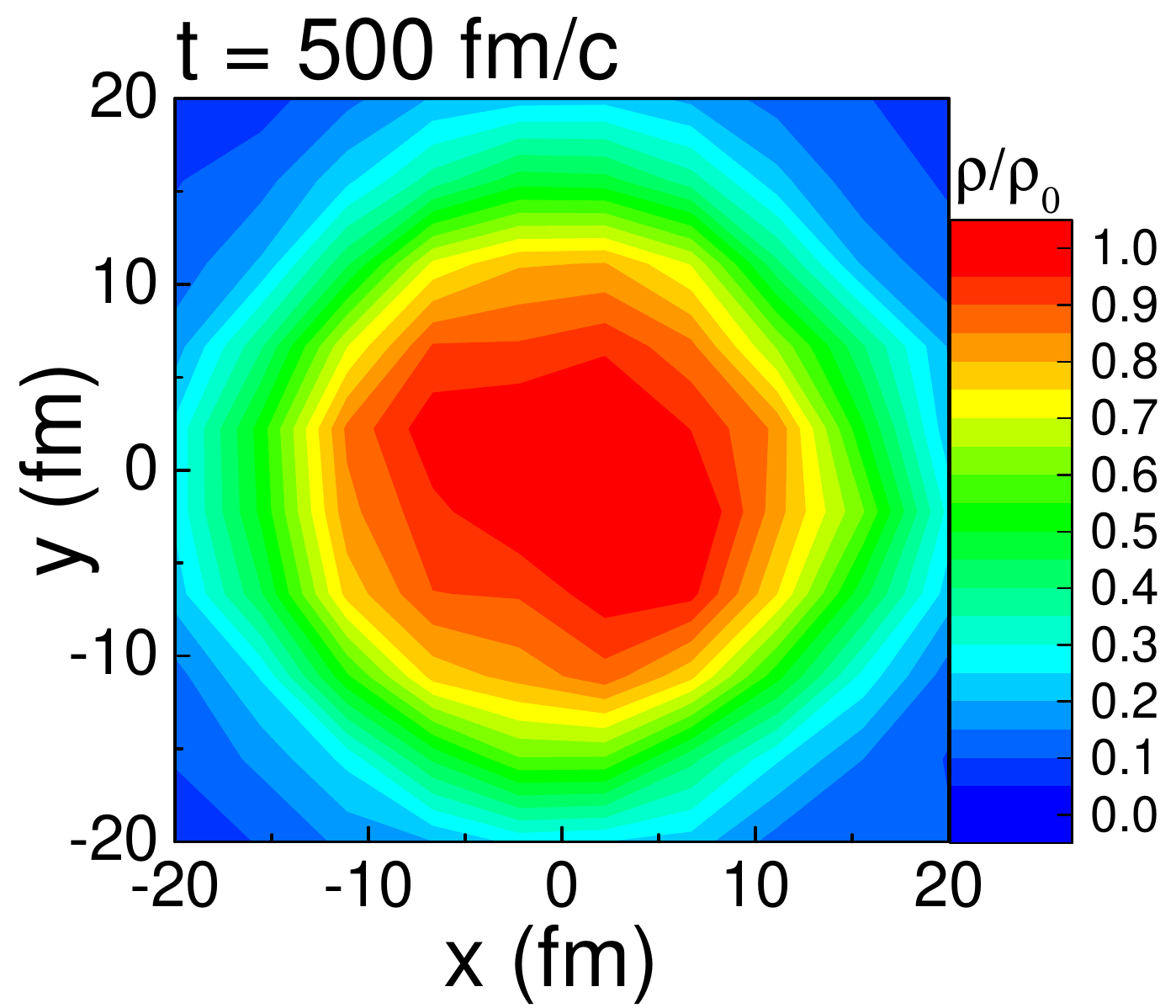}
\includegraphics[width=0.15\linewidth]{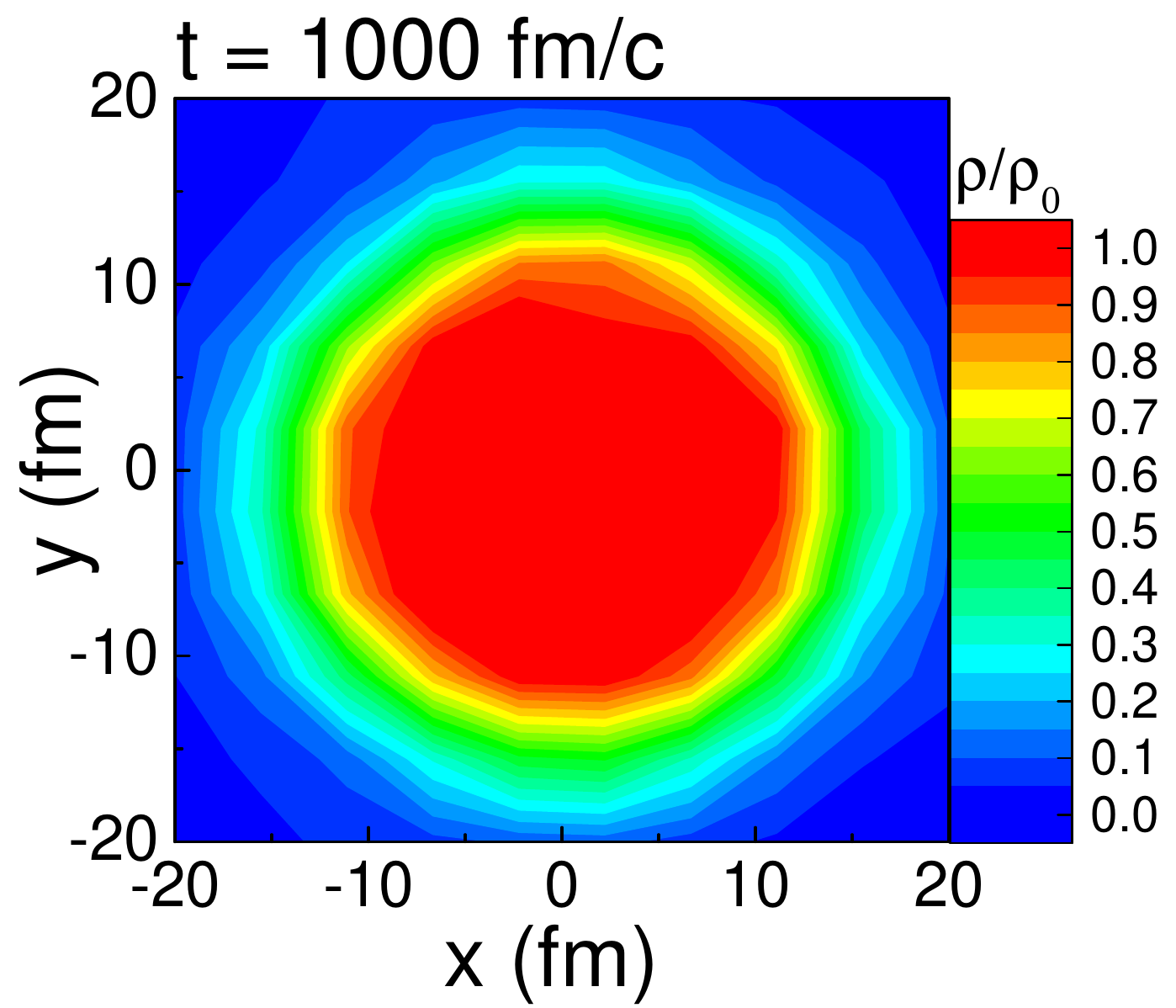}
\includegraphics[width=0.15\linewidth]{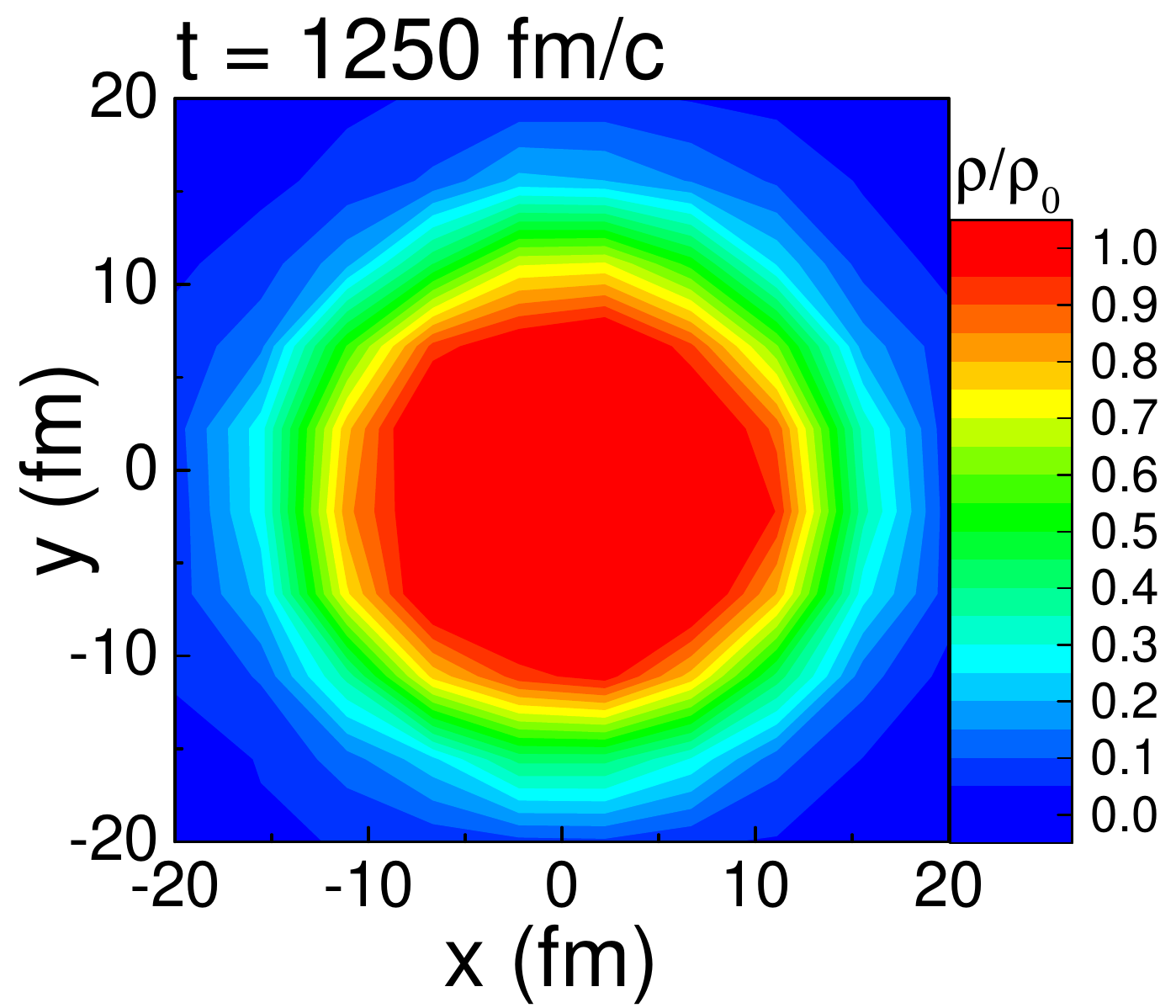}\\
\includegraphics[width=0.15\linewidth]{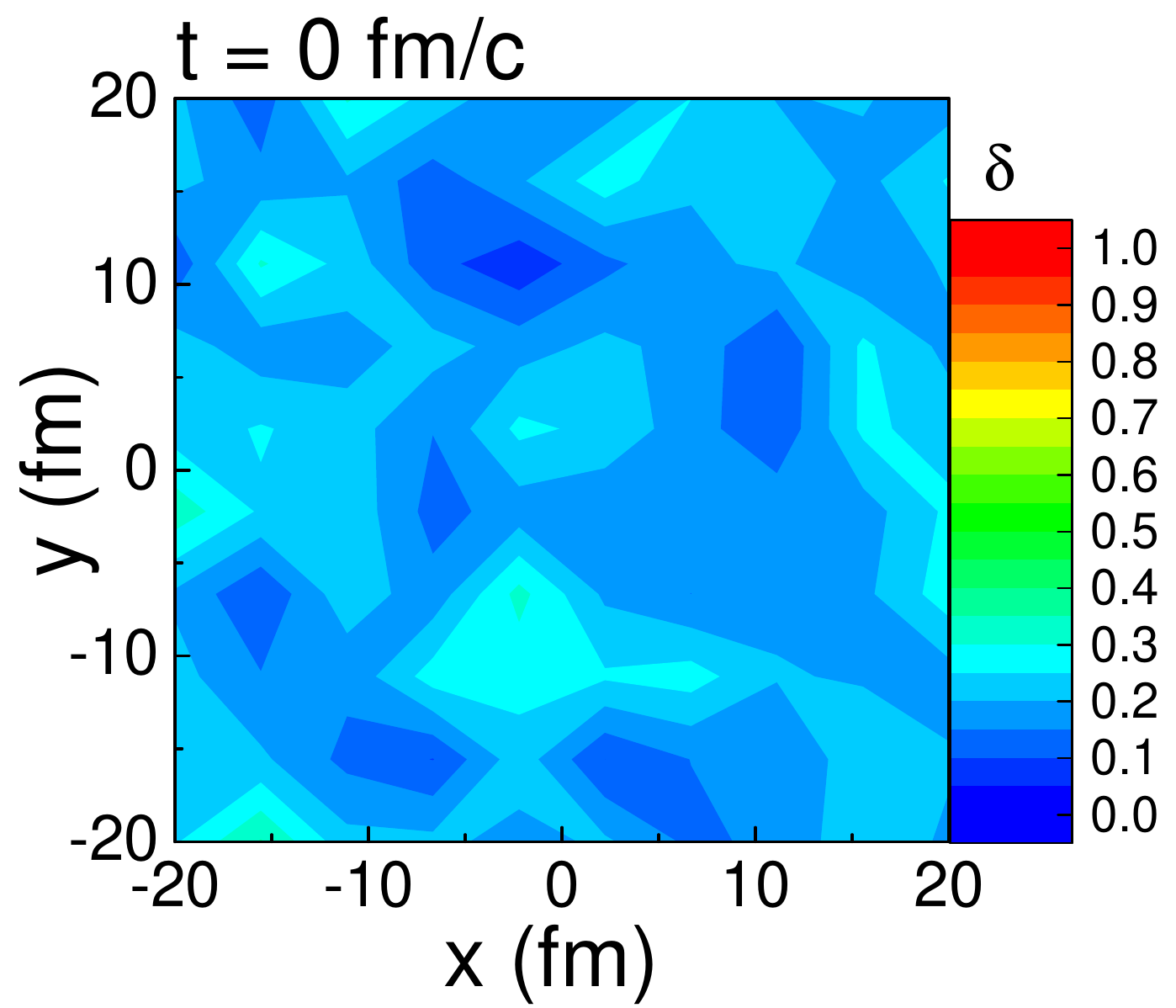}
\includegraphics[width=0.15\linewidth]{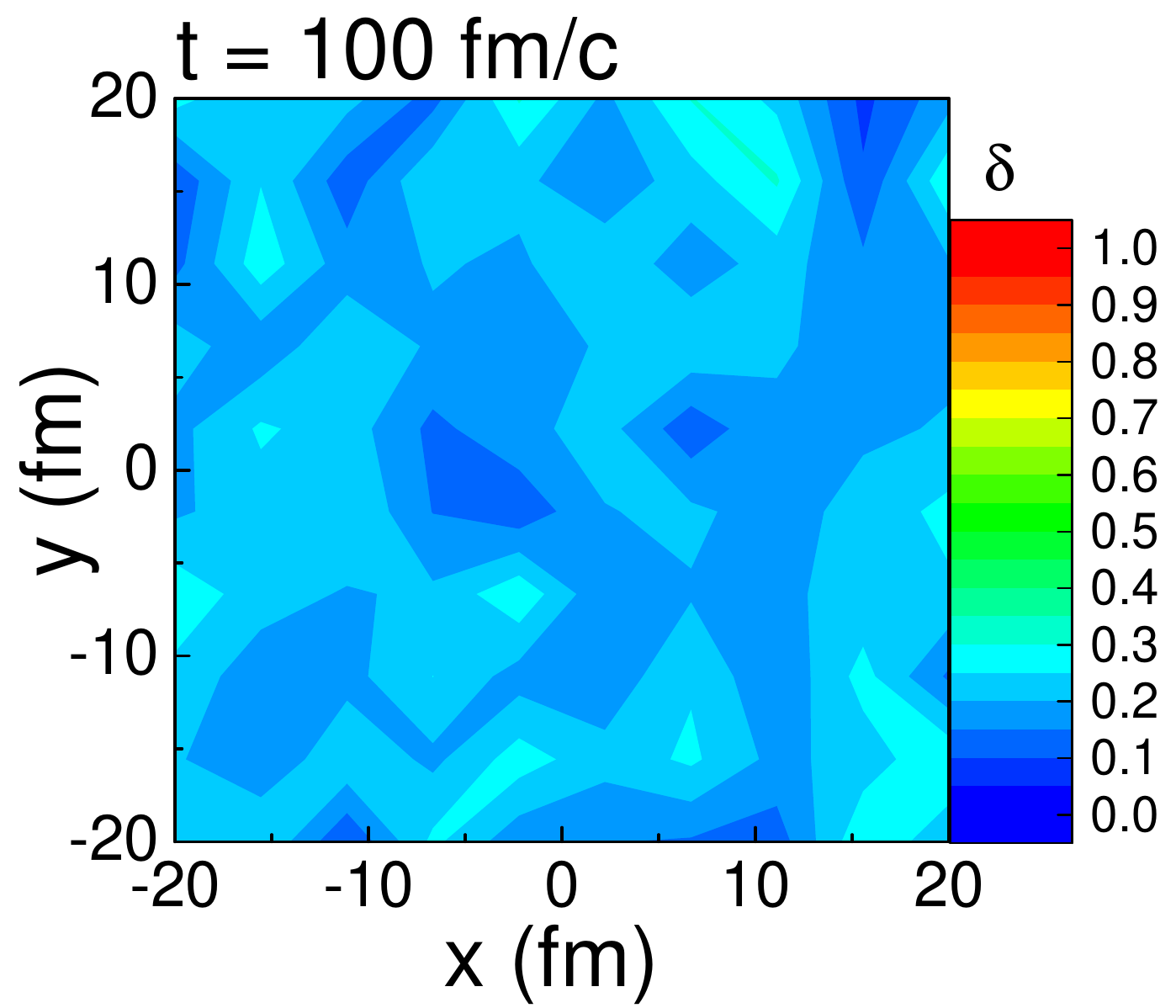}
\includegraphics[width=0.15\linewidth]{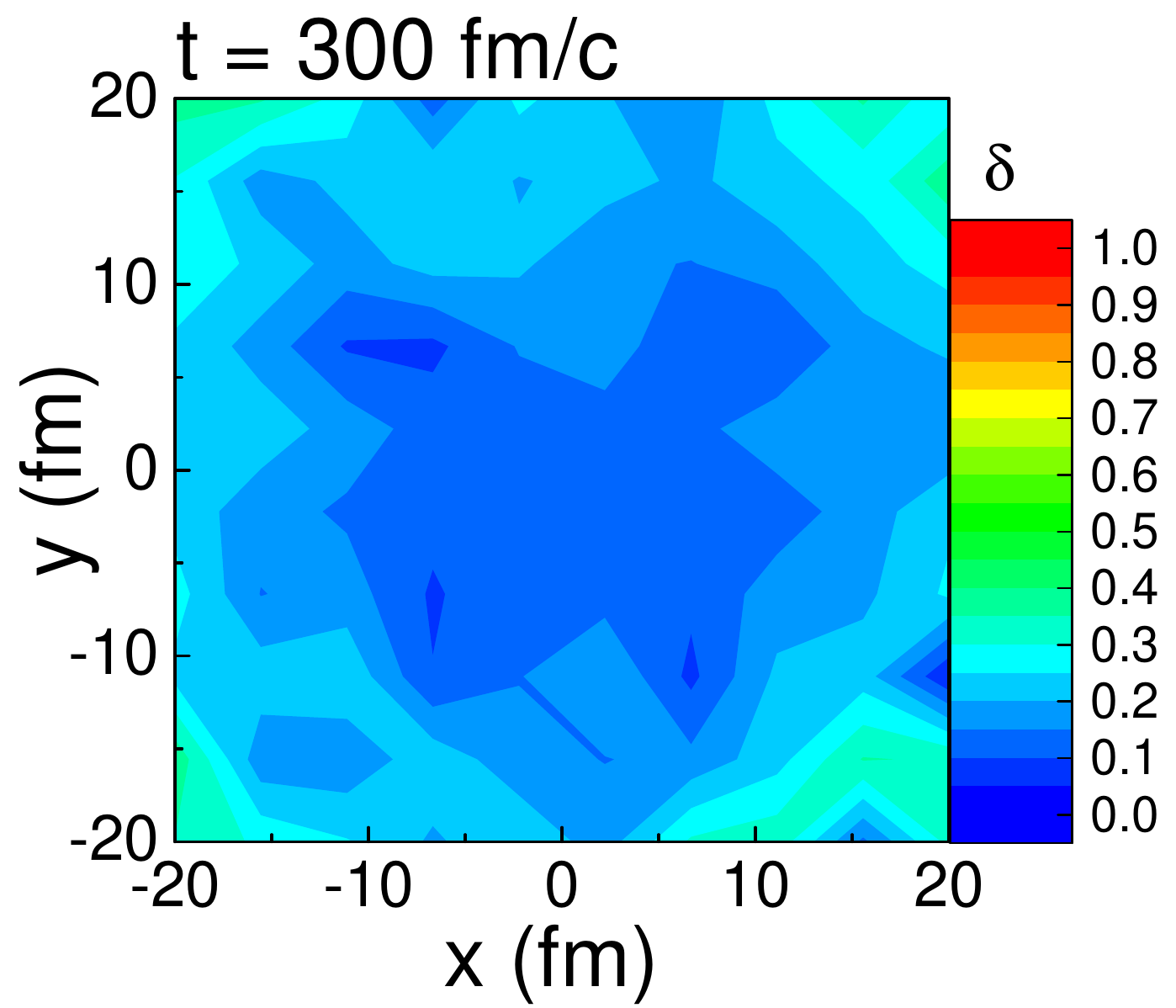}
\includegraphics[width=0.15\linewidth]{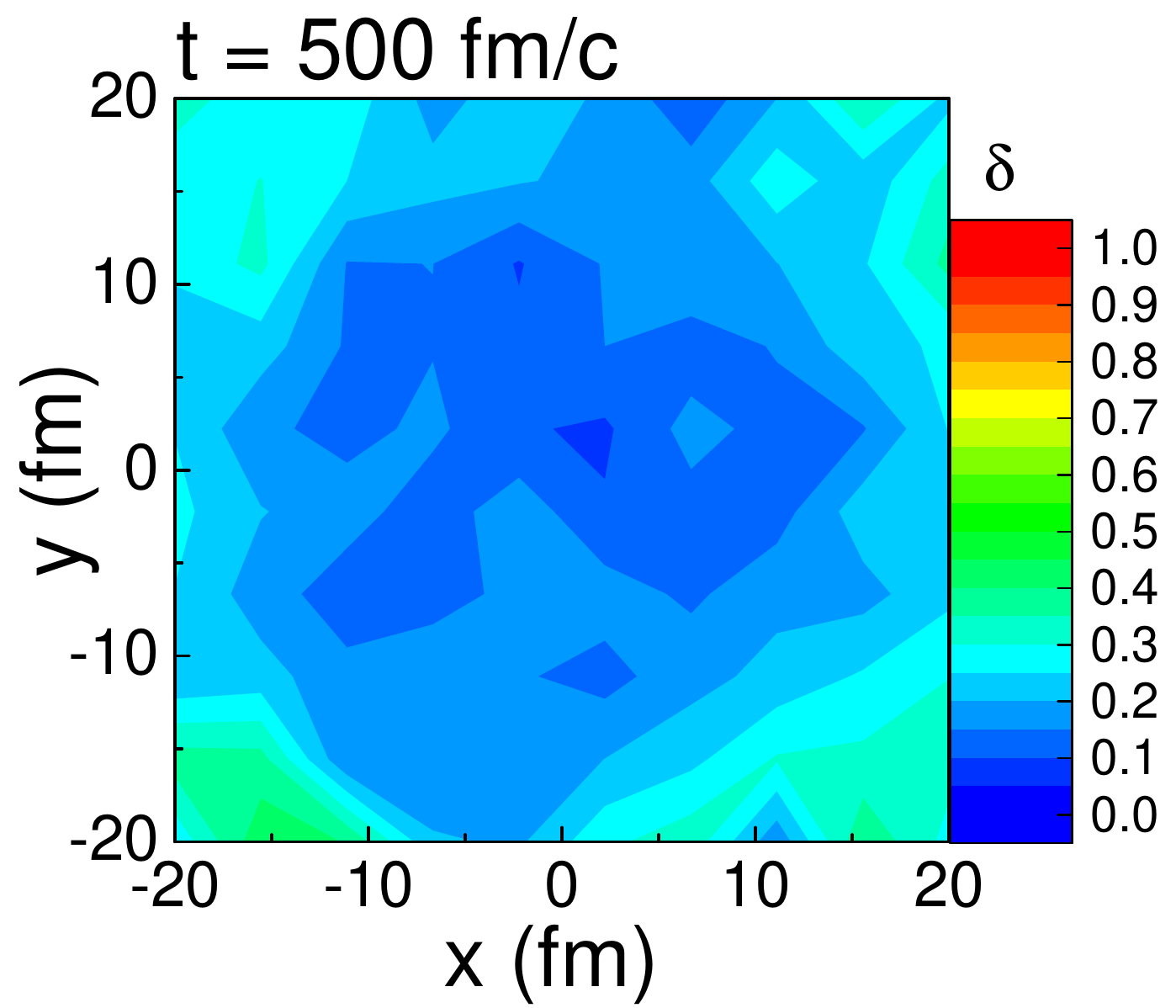}
\includegraphics[width=0.15\linewidth]{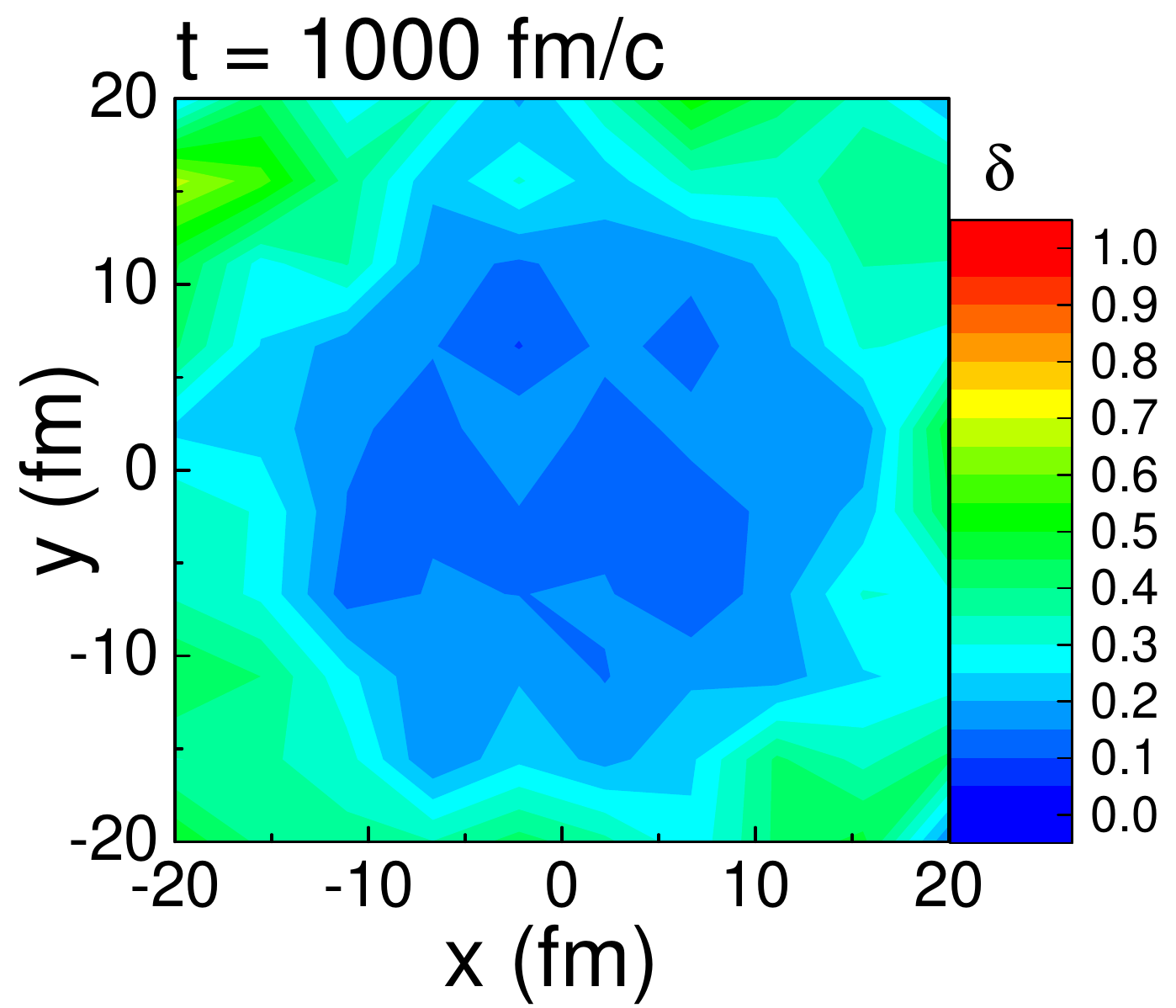}
\includegraphics[width=0.15\linewidth]{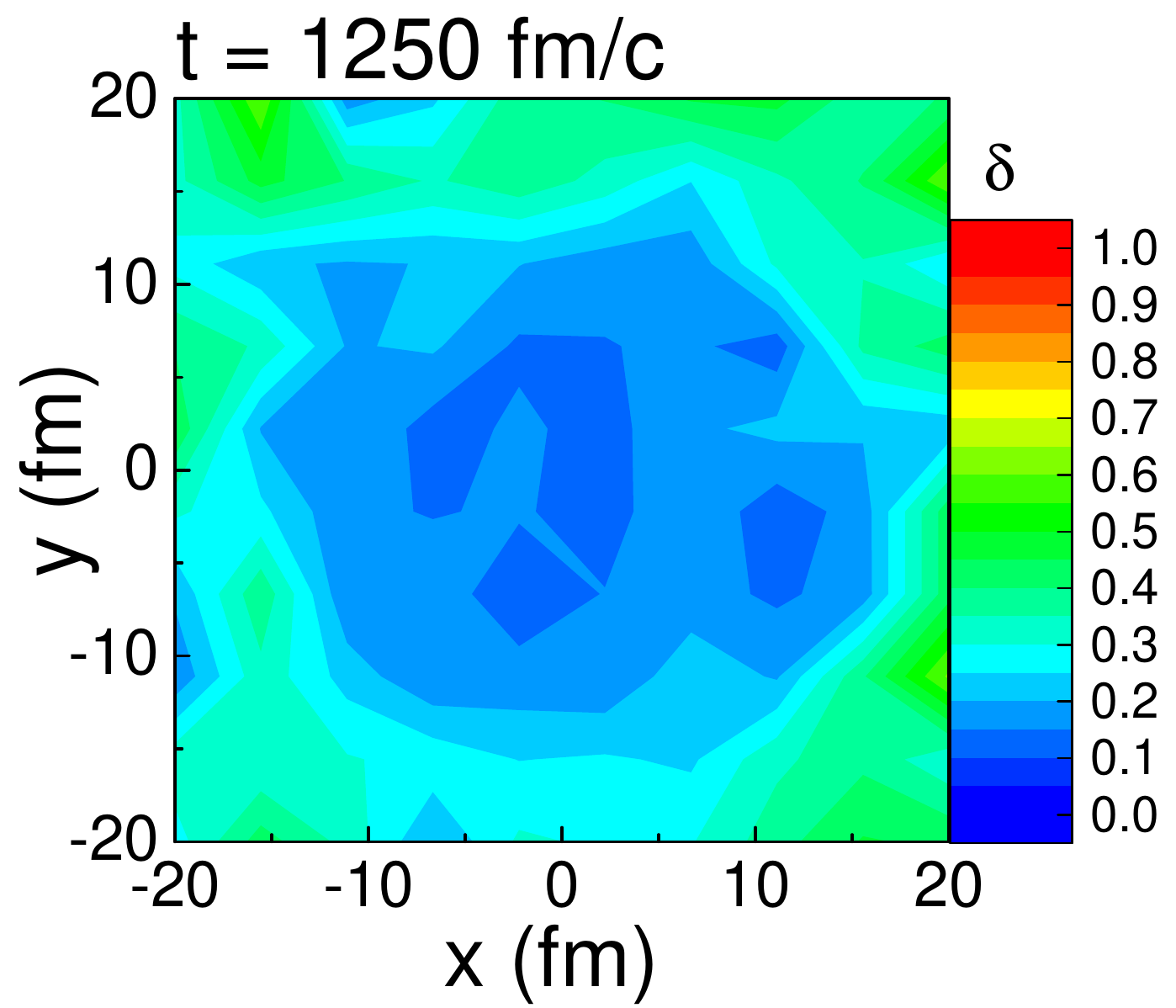}\\
\includegraphics[width=0.15\linewidth]{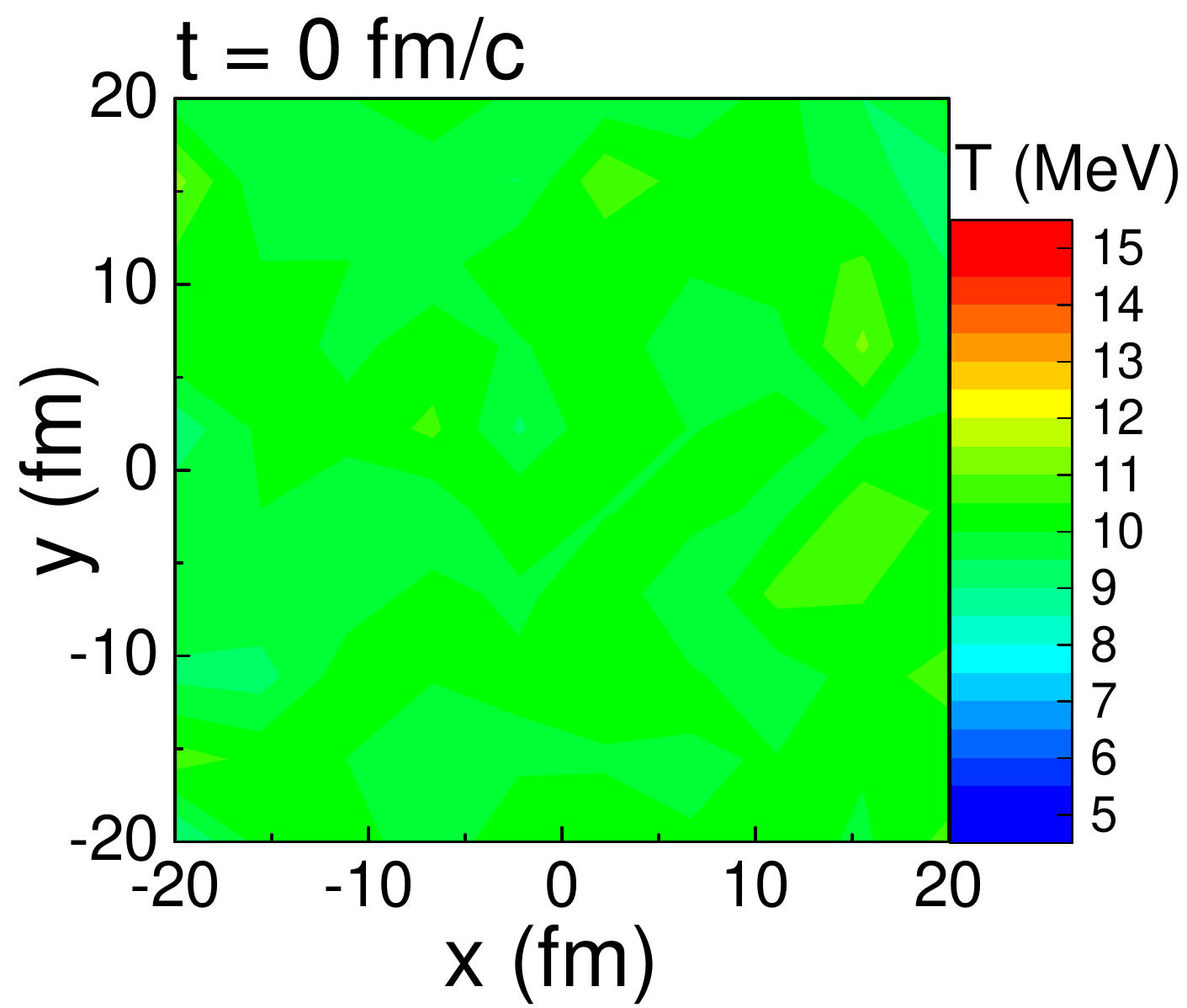}
\includegraphics[width=0.15\linewidth]{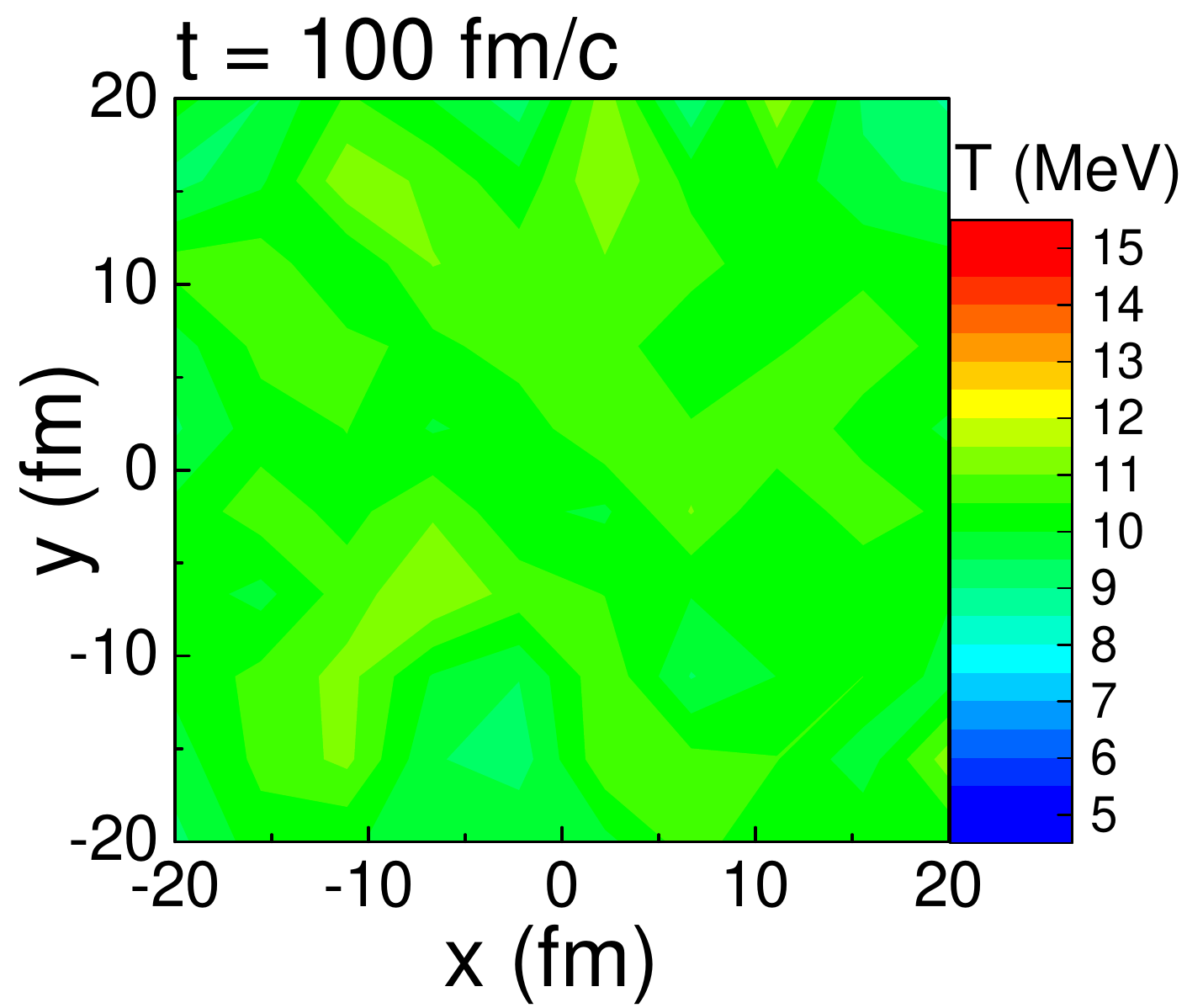}
\includegraphics[width=0.15\linewidth]{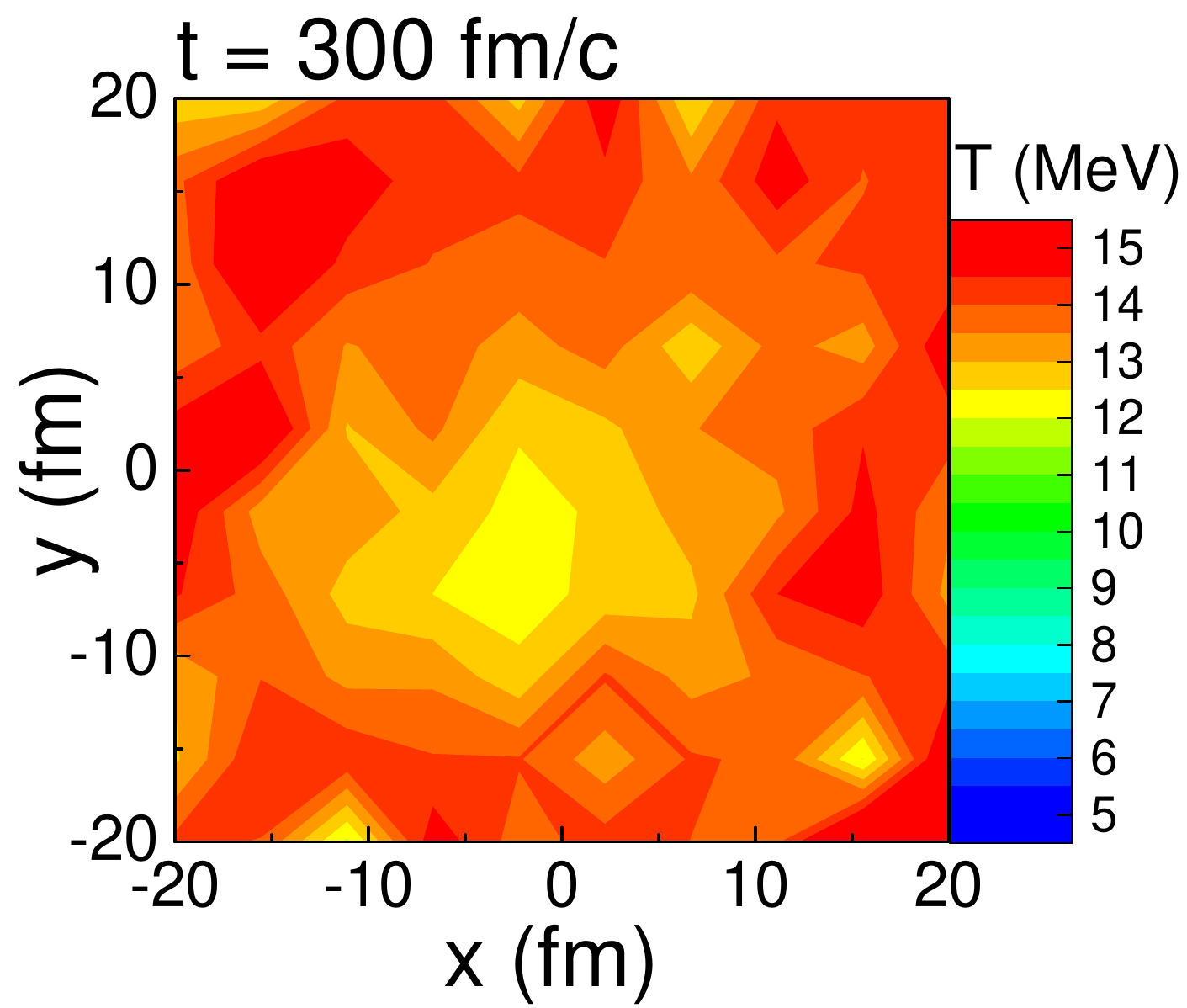}
\includegraphics[width=0.15\linewidth]{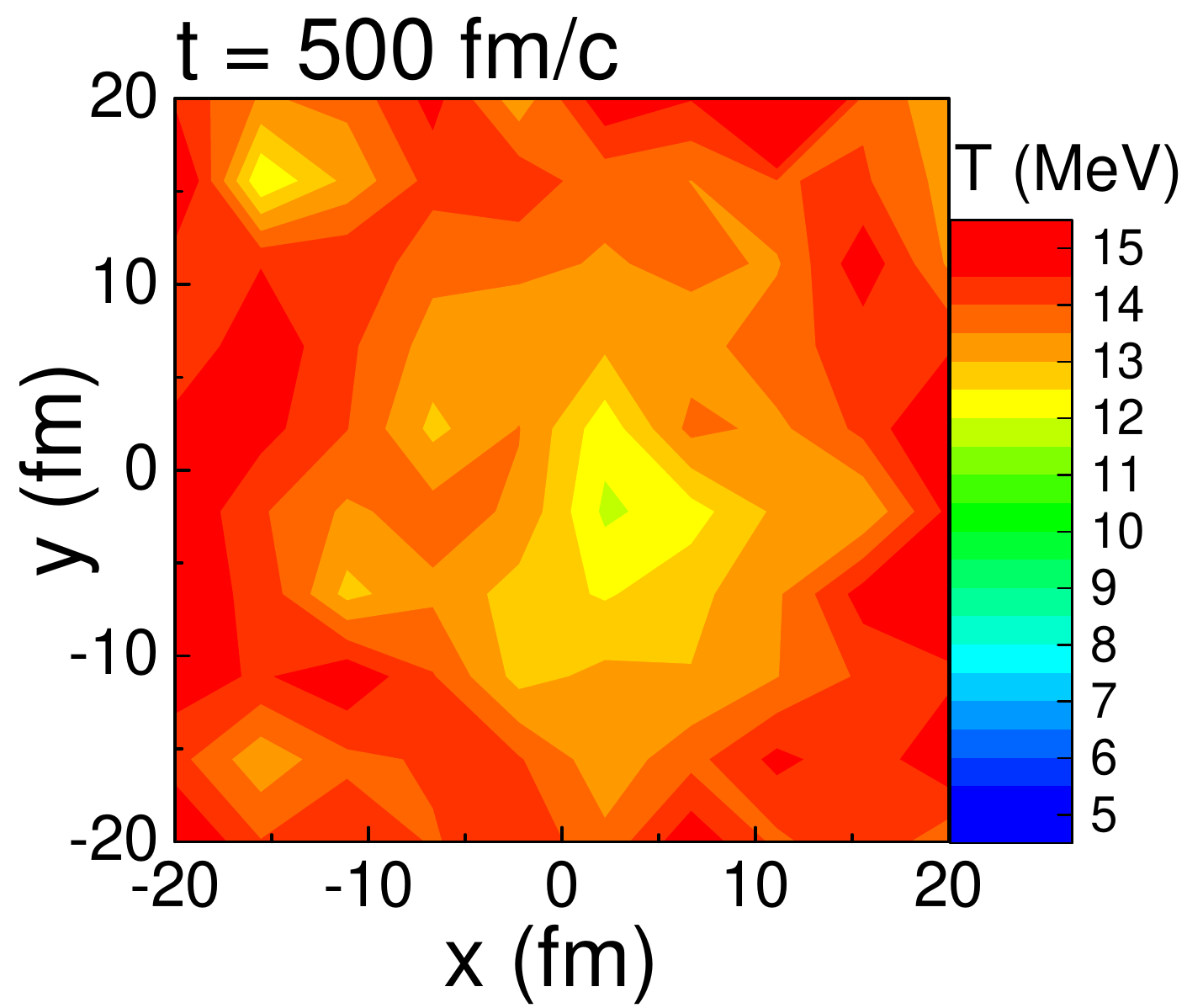}
\includegraphics[width=0.15\linewidth]{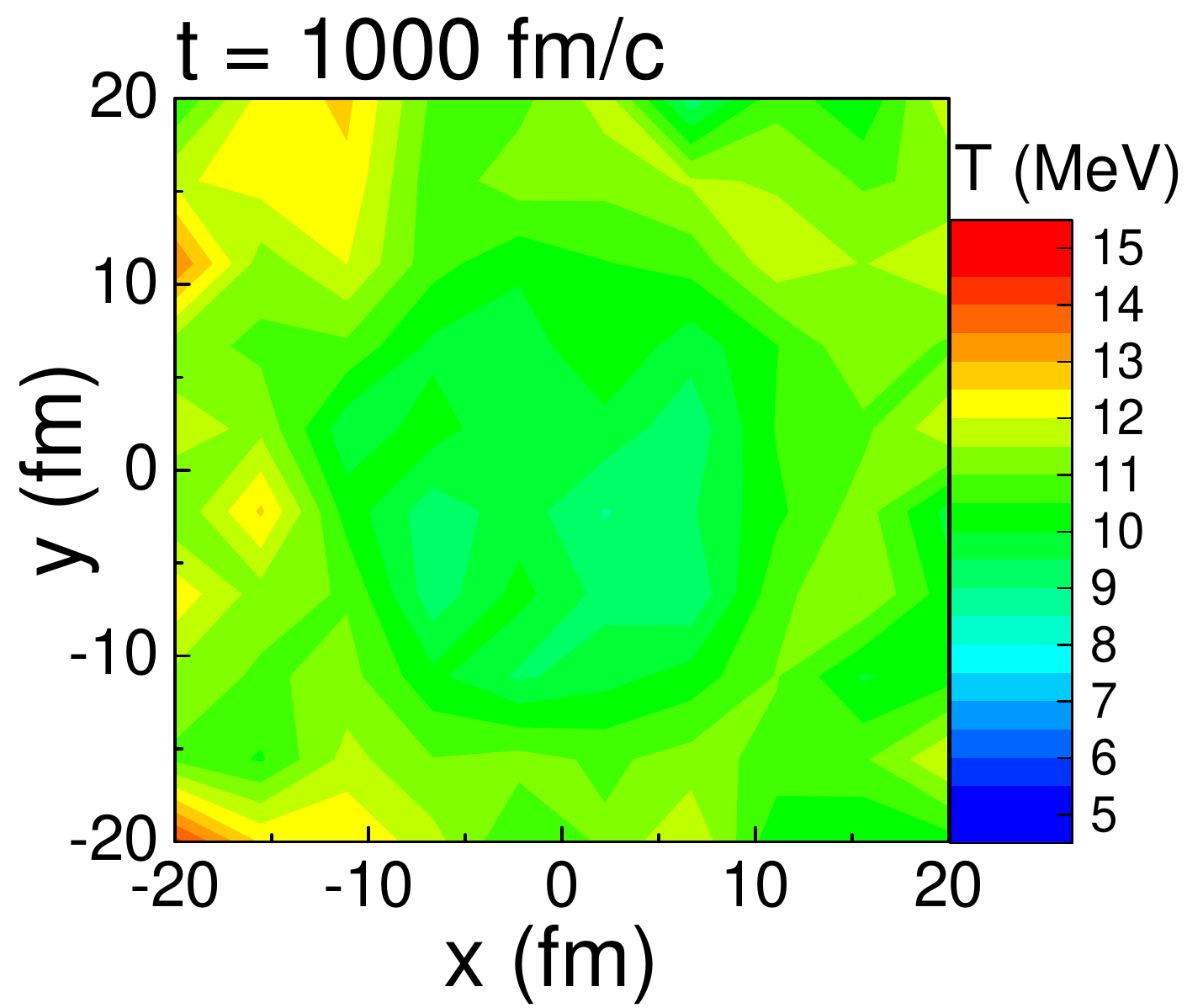}
\includegraphics[width=0.15\linewidth]{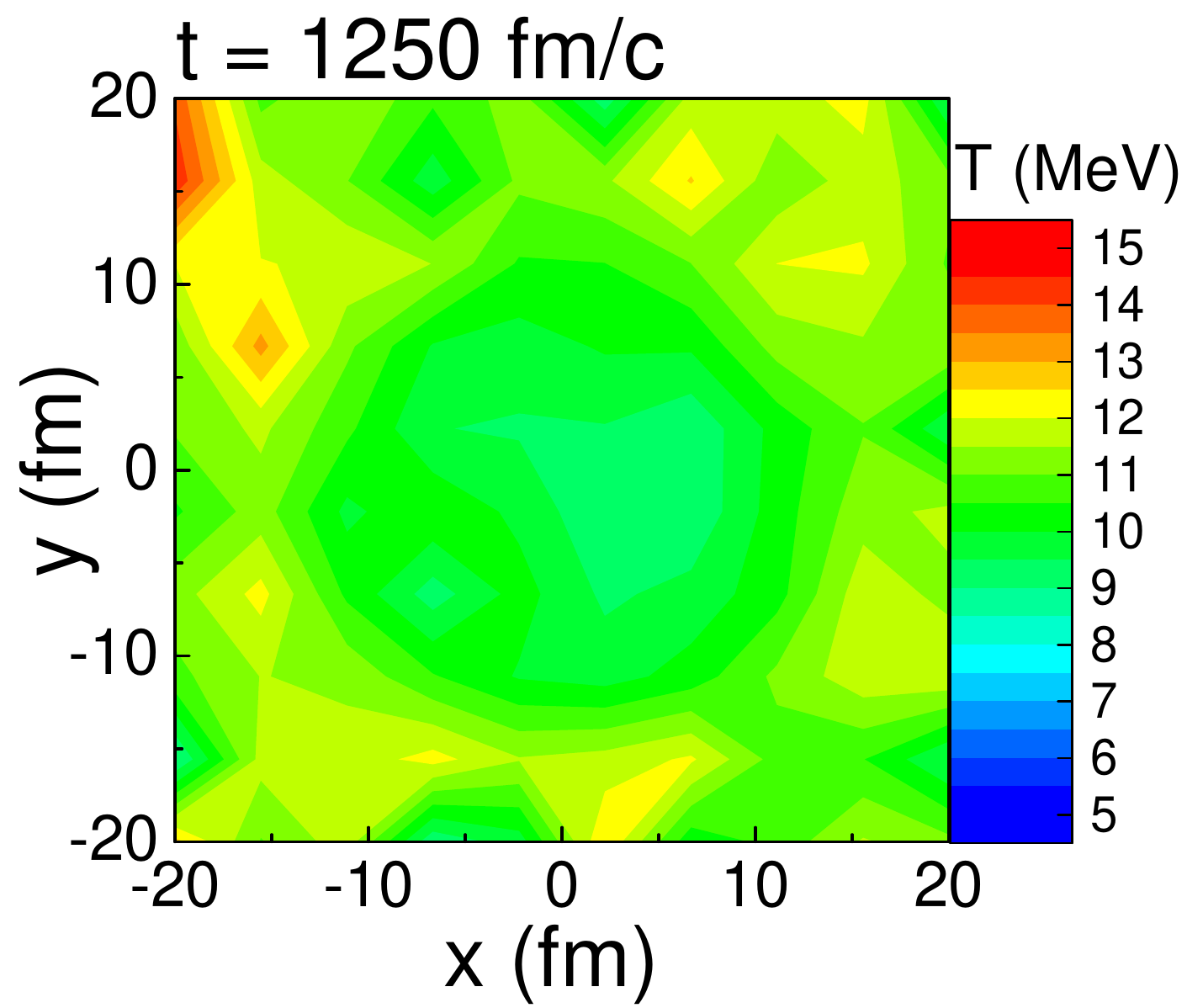}
\caption{\label{evo} Contours of the reduced density $\rho/\rho_0$ (first row), the isospin asymmetry $\delta$ (second row), and the temperature $T$ (third row) at time $t=0$, 100, 300, 500, 1000, and 1250 fm/c from IBUU simulations, starting from a uniform box system with initial density $\rho=0.3\rho_0$, isospin asymmetry $\delta = 0.2$, and temperature $T=10$ MeV, and with a reset of the temperature at $t=500$ fm/c.}
\end{figure*}

\begin{figure}[!h]
\includegraphics[width=0.6\linewidth]{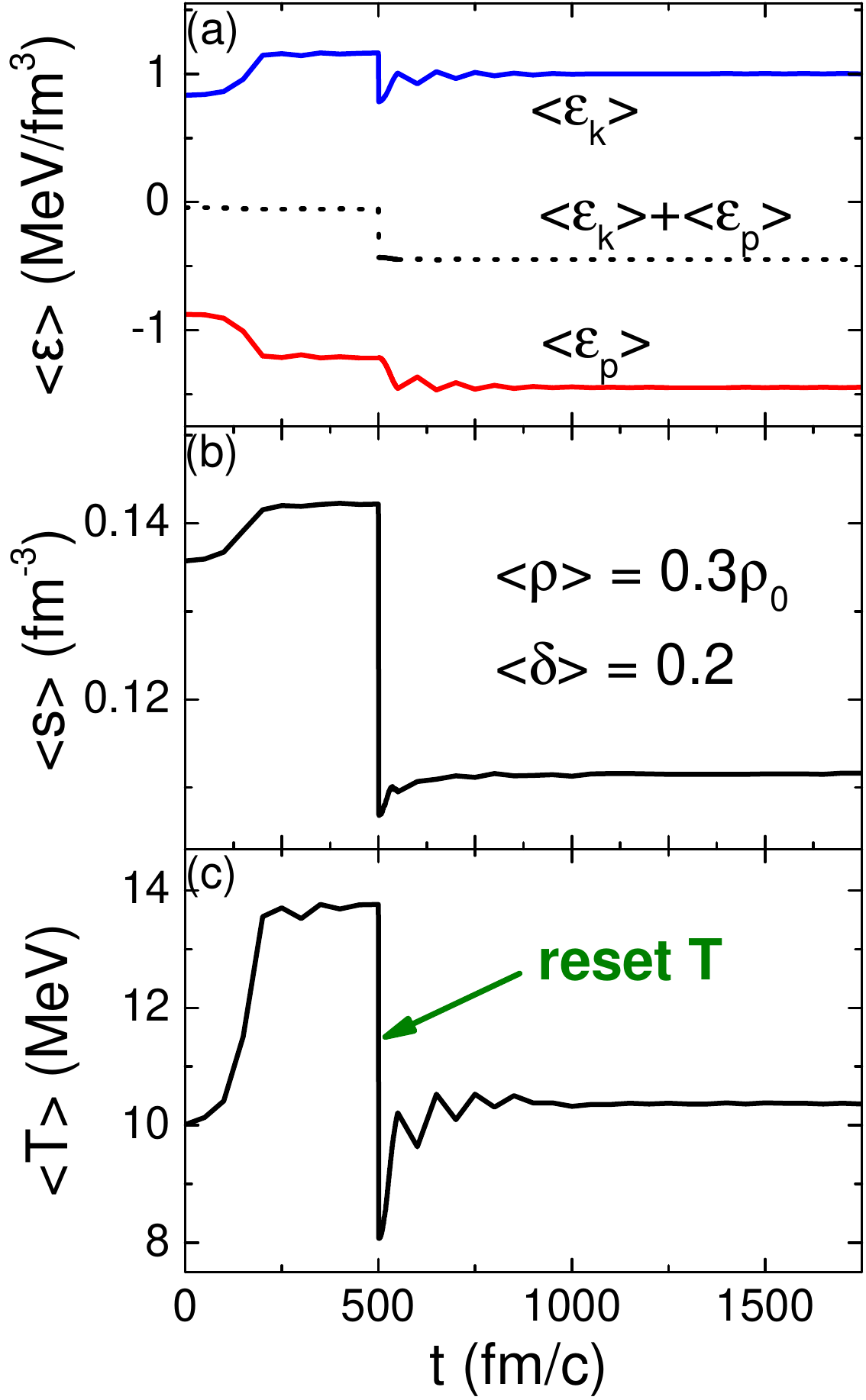}
\caption{\label{esT} Time evolution of physics quantities from IBUU simulations in a box system at an average density $\langle \rho \rangle=0.3\rho_0$, average isospin asymmetry $\langle \delta \rangle = 0.2$, and initial temperature $T=10$ MeV but with a reset of the temperature at $t=500$ fm/c. Top: Average kinetic energy density $\langle\epsilon_k\rangle$, potential energy density $\langle\epsilon_p\rangle$, and total energy density $\langle\epsilon_k\rangle+\langle\epsilon_p\rangle$; Middle: Average entropy density $\langle s \rangle$; Bottom: Average temperature $\langle T \rangle$. }
\end{figure}

In order to calculate the shear and bulk viscosities of nuclear matter in the spinodal region by using the Green-Kubo formulas, we need to prepare a dynamically equilibrated box system with nuclear clusters. The method is described as follows and illustrated in Figs.~\ref{evo} and \ref{esT}. For example, we start from a uniform system with density $\rho=0.3\rho_0$, isospin asymmetry $\delta=0.2$, and temperature $T=10$ MeV. Since the state of the initial nuclear matter is in the spinodal region, as shown in Fig.~\ref{pd}, the uniform matter gradually evolves to the mixture of higher-density clusters and lower-density nucleon gas, as can be shown from the contours at $t<500$ fm/c in Fig.~\ref{evo}. It is seen that the higher-density (lower-density) liquid (gas) phase has a smaller (larger) isospin asymmetry, consistent with the thermal calculation of the liquid-gas mixed phase~\cite{Xu:2007eq}. Since the system with a mixture of the liquid and gas phase has a lower potential energy compared to a uniform one at the same average density and isospin asymmetry, the kinetic energy and the overall temperature increase compared to the initial state, and this can be seen in Fig.~\ref{esT}. This becomes troublesome if we want to generate dynamically equilibrated systems at low temperatures. To overcome this problem, we take the system at $t=500$ fm/c as a new initial state and reset the temperature of the system by resampling the nucleon momentum distribution in each local cell. In this case, there is a sudden decrease of the average kinetic energy density $\langle \epsilon_k \rangle$, the average total energy density $\langle \epsilon \rangle$, the average entropy density $\langle s \rangle$, and the average temperature $\langle T \rangle$, though the average potential energy density $\langle \epsilon_p \rangle$ does not have an instant change. After a short relaxation time, the system evolves to a new equilibrated state, with a higher temperature similar to the initial one. This is also shown by contours in Fig.~\ref{evo}, where one can see that the density and the isospin asymmetry actually do not change by much at $t>500$ fm/c. After $t=1250$ fm/c, the system has reached a dynamically equilibrated state with nuclear clusters. We thus use $t_0=1250$ fm/c, the value of which can be slightly different for different initial states of nuclear systems, as the starting time to calculate the shear and bulk viscosities through Eqs.~(\ref{eta0}) and (\ref{zeta0}). Since in non-uniform systems we mostly talk about average quantities over space, such average symbol ``$\langle...\rangle$'' will be omitted in subsequent discussions.

\subsection{shear and bulk viscosities}
\label{c}

\begin{figure}[!h]
\includegraphics[width=1.0\linewidth]{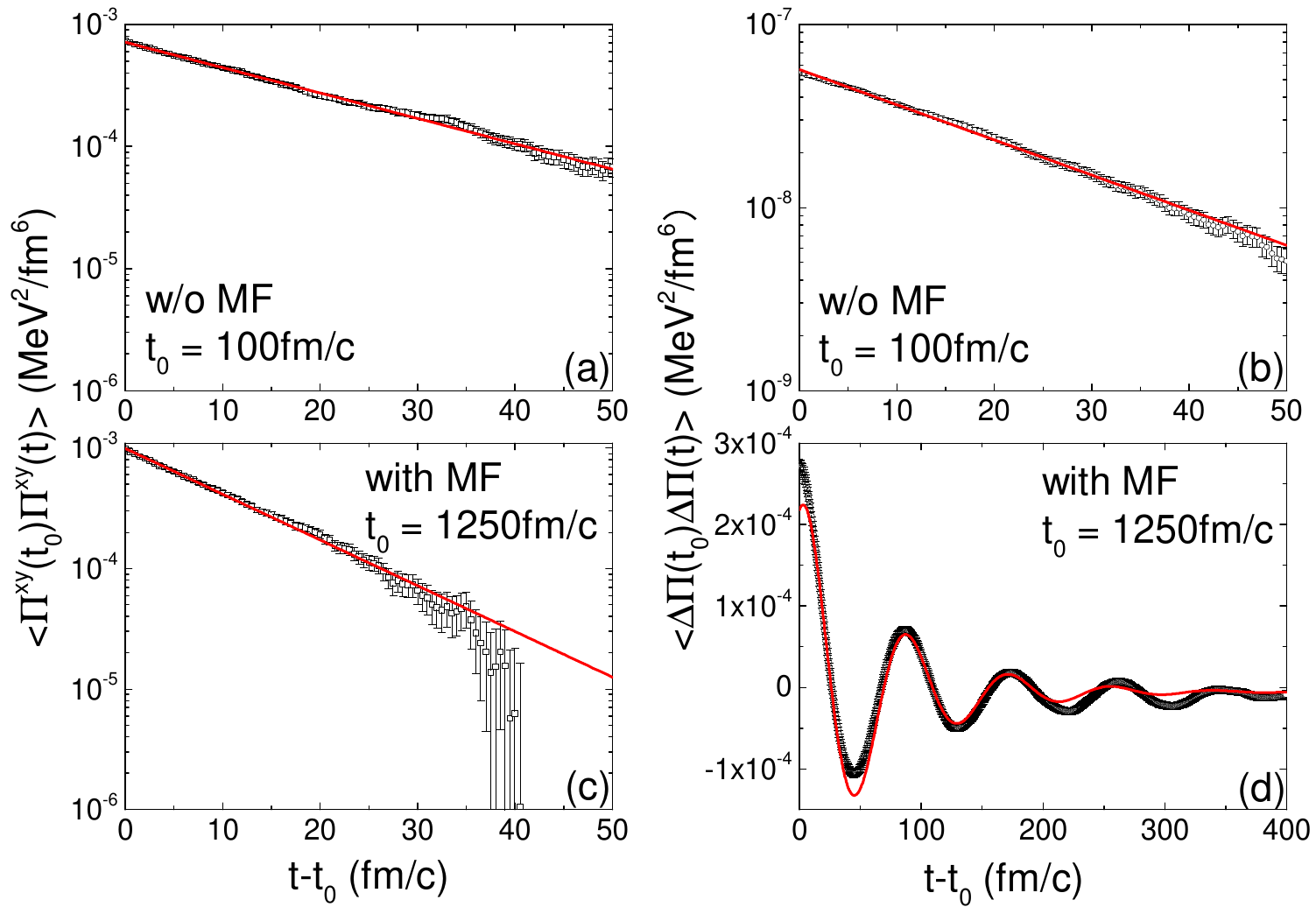}
\caption{\label{fit} Time dependence of correlations of the shear (left) and bulk (right) components of the energy-momentum tensor, for a nuclear system with an average density $\rho=0.3\rho_0$, average isospin asymmetry $\delta=0.2$, and average temperature of about 10 MeV. Results for a uniform system without mean-field potential (MF) are shown in upper panels, and those for a non-uniform system with mean-field potential are shown in lower panels.}
\end{figure}

\begin{figure}[!h]
\includegraphics[width=1.0\linewidth]{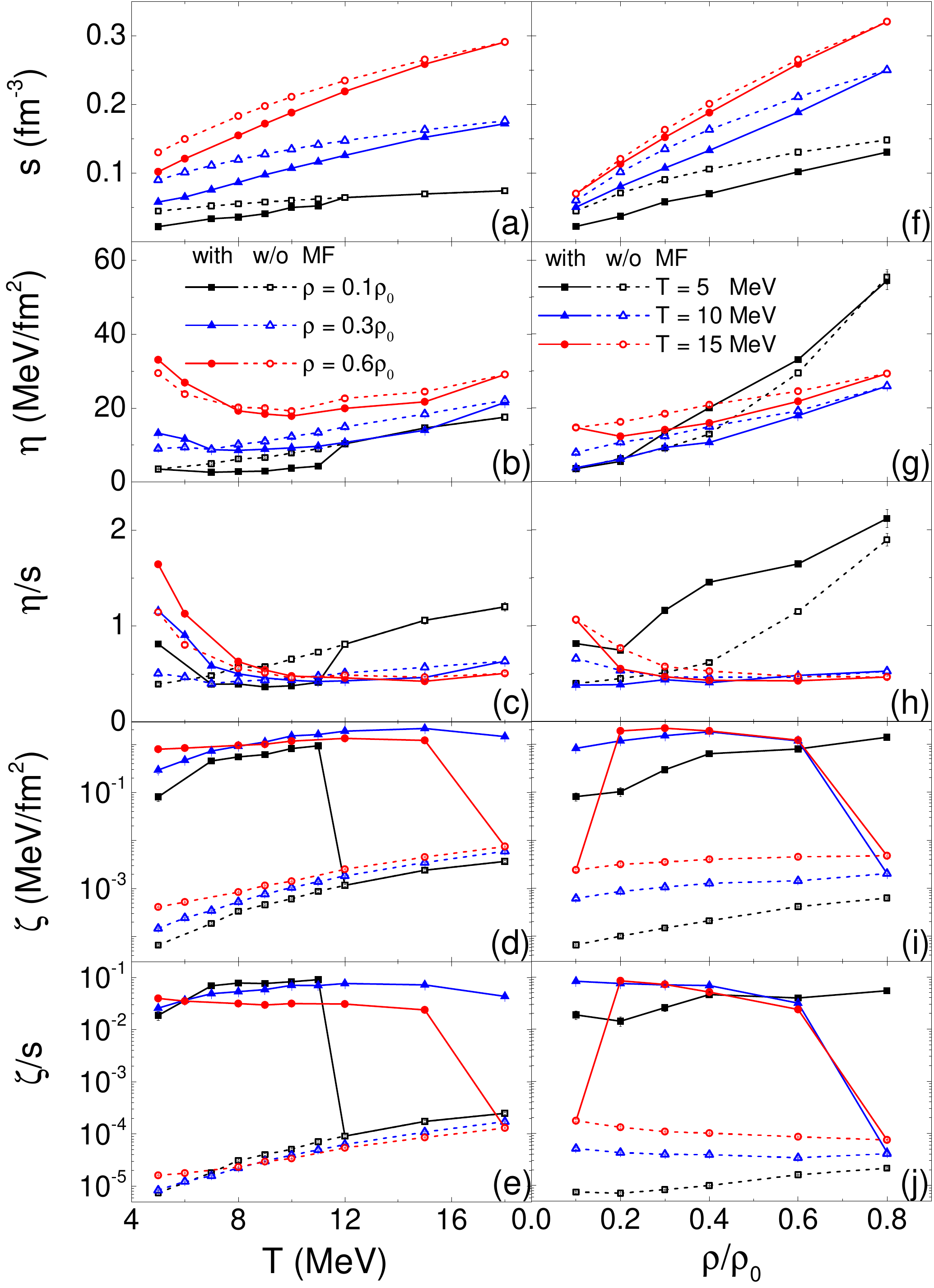}
\caption{\label{Trho} Left: Temperature dependence of the entropy density ($s$), the shear viscosity ($\eta$), the specific shear viscosity ($\eta/s$), the bulk viscosity ($\zeta$), and the specific bulk viscosity ($\zeta/s$) in isospin symmetric nuclear matter at fixed average densities $\rho=0.1$, 0.3, and $0.6\rho_0$; Right: Density dependence of these quantities at fixed average temperatures $T=5$, 10, 15 MeV. Results with and without mean-field potential (MF) are compared.}
\end{figure}

The time evolutions of correlations of the shear and bulk components of the energy-momentum tensor after the system has reached equilibrium are displayed in Fig.~\ref{fit}. In the case without mean-field potential, the system remains uniform even if the initial state is in the spinodal region, and a relaxation time of $100$ fm/c is used, as shown in the upper panels of Fig.~\ref{fit}. It is seen that the correlations for both the shear and the bulk components can be well fitted by an exponential decay function so that the integral over time can be calculated as
\begin{equation}
\label{exp1}
\int_{t_0}^\infty Ae^{-B(t-t_0)} dt = \frac{A}{B}.
\end{equation}
With mean-field potential, the dynamically equilibrated system is prepared using the method as described in Figs.~\ref{evo} and \ref{esT}. Compared to the uniform system, the correlation of the shear component of the energy-momentum tensor has a larger initial value at $t=t_0$ due to the enhanced correlation in the presence of nuclear clusters, while the correlation decays more quickly due to more successful collisions in hot clusters. This leads to both larger $A$ and larger $B$ in Eq.~(\ref{exp1}), and whether the resulting shear viscosity $\eta$ is larger or smaller in non-uniform nuclear matter depends on such competition effect. The time evolution of the correlation of the bulk component of the energy-momentum tensor has a completely different behavior in non-uniform nuclear system, and this is likely due to the collective oscillation of the clusters similar to the giant monopole resonance. Such behavior can be fitted by a sinusoidal function with its magnitude decreasing exponentially with time, so the integral over time can be calculated as
\begin{eqnarray}
\label{exp2}
&&\int_{t_0}^\infty Ae^{-B(t-t_0)}\sin[C(t-t_0)+D] dt \notag\\
&=& \frac{AB\sin(D)+AC\cos(D)}{B^2+C^2}.
\end{eqnarray}
The equilibrium value $\Pi_{eq}$ of the bulk component of the energy-momentum tensor is carefully chosen so that $\langle \Delta \Pi(t_0) \Delta \Pi(t)\rangle$ is nearly 0 for $t \rightarrow \infty$. While the oscillation frequency and the initial phase shift characterized respectively by $C$ and $D$ may have minor effects on the results, the bulk viscosity is seen to be mostly determined by $A$ and $B$. Compared to the case of uniform system, while the decay trend characterized by $B$ is also larger due to more successful collisions, the initial value of the correlation characterized by $A$ is significantly larger due to the enhanced correlation with clusters. This leads to a significantly larger bulk viscosity $\zeta$ in systems with nuclear clusters, compared to that in uniform nuclear matter without mean-field potential. Here we emphasize again that a greater enhancement of the initial $\left\langle \Delta \Pi(t_0) \Delta \Pi(t)\right\rangle$ than the initial $\left\langle\Pi^{x y}(t_0) \Pi^{x y}(t)\right\rangle$ in non-uniform matter compared to that in uniform matter is observed, due to the redistribution of the bulk component of the energy-momentum tensor with a constrained total energy.

With the method described in Fig.~\ref{fit}, we calculate extensively the shear and bulk viscosities at different densities and temperatures for isospin symmetric nuclear matter, and these results are displayed in Fig.~\ref{Trho}. At the states that the systems are out of the spinodal region and keep uniform even with mean-field potential, the results are identical to those in systems without mean-field potential. Consistent with Ref.~\cite{Hua:2023amo}, the entropy density $s$, which increases with increasing temperature or density, is generally larger in a uniform system without mean-field potential, compared to that in a non-uniform system. For the shear viscosity $\eta$, while it is generally smaller with clusters for a constant and isotropic NN cross section as seen in Ref.~\cite{Hua:2023amo}, some new features are observed with more realistic energy-dependent and anisotropic cross sections. Here, while $\eta$ is generally smaller in non-uniform systems than in uniform systems at higher temperatures, it is smaller in uniform systems than in non-uniform systems at lower temperatures. The former is due to the enhanced collisions in the presence of nuclear clusters, as mentioned in Ref.~\cite{Hua:2023amo}. The latter is due to significantly more successful collisions as a result of large NN cross sections at lower kinetic energies as shown in Fig.~\ref{cs}, compared to that from a constant NN cross section of 40 mb used in Ref.~\cite{Hua:2023amo}. This leads to a stronger decay of the correlation function (see Fig.~\ref{fit} (c) and Eq.~(\ref{exp1})), and such effect is larger in uniform than in non-uniform matter. While $\eta$ generally increases with increasing density, the energy-dependent NN cross sections make $\eta$ decrease (increase) with increasing temperature at higher (lower) densities, regardless of the clustering effect. Thus, while the temperature dependence of the specific shear viscosity $\eta/s$ may have a minimum point, the temperature of the minimum $\eta/s$ is different from that for a constant NN cross section in Ref.~\cite{Hua:2023amo}. For the bulk viscosity $\zeta$, it generally increases with the increasing temperature and/or density, due to larger bulk components of the energy-momentum tensor at higher temperatures/densities. On the other hand, it is remarkable to see that $\zeta$ is $2-3$ orders of magnitude larger in non-uniform systems, due to the stronger correlation of the bulk components of the energy-momentum tensor induced by nuclear clustering as shown in Fig.~\ref{fit} (d) and described by Eq.~(\ref{exp2}). In such cases, there are complicated competition effects between the correlation of the energy-momentum tensor from nuclear clustering and NN collisions, as anticipated in the discussion of Eq.~(\ref{exp2}). Overall, the larger $\zeta/s$ with clusters compared to that in uniform matter is thus insensitive to the energy dependence of NN cross sections.

\begin{figure}[!h]
\includegraphics[width=1.0\linewidth]{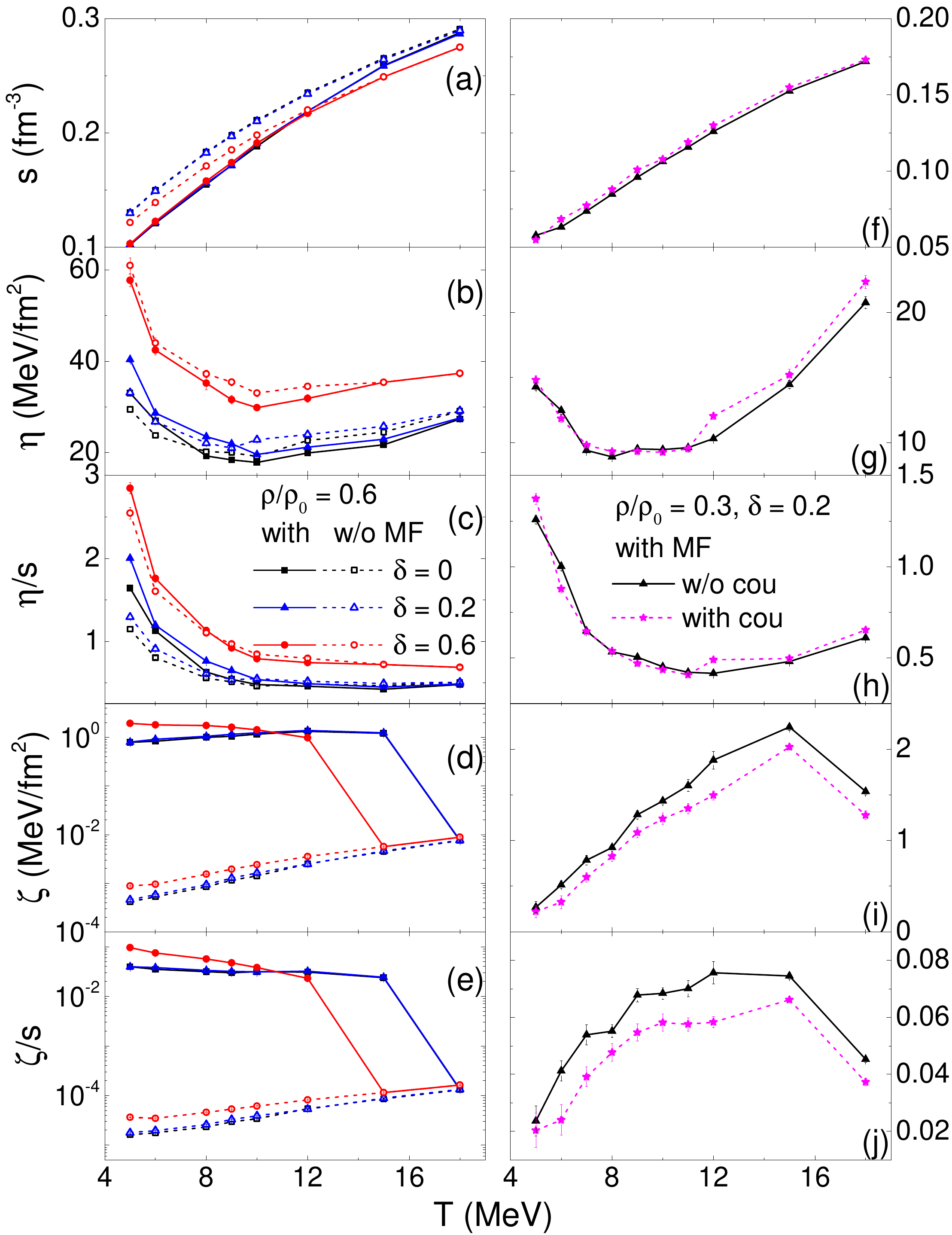}
\caption{\label{deltacou} Left: Temperature dependence of the entropy density ($s$), the shear viscosity ($\eta$), the specific shear viscosity ($\eta/s$), the bulk viscosity ($\zeta$), and the specific bulk viscosity ($\zeta/s$) at a fixed average density $\rho =0.6\rho_0$ in isospin symmetric ($\delta=0$) and asymmetric ($\delta=0.2$ and 0.6) nuclear matter with and without mean-field potential (MF); Right: Temperature dependence of these quantities at a fixed average density $\rho=0.3 \rho_0$ and isospin asymmetry $\delta=0.2$ with and without Coulomb potential (cou) in the presence of mean-field potential.}
\end{figure}

We have further investigated the isospin dependence of these quantities, and the results in nuclear matter at different average isospin asymmetries are compared in the left panels of Fig.~\ref{deltacou}. Since the isospin dependence has been found to be weaker at lower densities, here we choose a relatively higher average density $\rho=0.6\rho_0$. It is seen that the isospin asymmetry of the system has minor effects on the entropy density. At $\delta=0.2$, both $\eta$ and $\zeta$ as well as the corresponding specific viscosities have a qualitatively similar behavior compared to that at $\delta=0$, while they are all quantitatively larger. The latter can be traced back to the smaller neutron-neutron than neutron-proton cross section as shown in Fig.~\ref{cs} (a), which leads to less successful collisions in asymmetric matter compared to those in symmetric matter, thus leading to weaker decays of the correlations in Eqs.~(\ref{exp1}) and (\ref{exp2}). While the spinodal region is similar for $\delta=0$ and 0.2, it shrinks dramatically for $\delta=0.6$, as can be seen in Fig.~\ref{pd}. Consequently, the values of $\zeta$ and $\zeta/s$, which are more sensitive to nuclear clustering, drop at a lower temperature compared to the case for a smaller $\delta$. While it is impossible to reach such a high isospin asymmetry in heavy-ion collisions, such system could exist in neutron-star mergers.

While the Coulomb interaction has been neglected in all above calculations, we devote the right panels of Fig.~\ref{deltacou} to illustrate the Coulomb effect on viscosities in nuclear matter at an average density $\rho=0.3 \rho_0$ and isospin asymmetry $\delta=0.2$. Here, the Coulomb potential between protons is incorporated based on the lattice Hamiltonian framework as described in Ref.~\cite{Hua:2023amo}. Basically, the repulsive Coulomb interaction weakens the clustering effect. This leads to a slight increase of $s$ as well as $\eta$ and $\eta/s$ at higher temperatures, and a considerable decrease of $\zeta$ and $\zeta/s$, and these effects are consistent with the discussions in previous paragraphs. The behaviors of these quantities are seen to remain qualitatively similar with or without Coulomb potential.

\section{conclusions}

We have studied the shear and bulk viscosities in the spinodal region of isospin asymmetric nuclear matter by using transport simulations in a box system, with calibrated mean-field calculation and NN collisions. The Green-Kubo method is used to calculate the viscosities in the prepared dynamically equilibrated system with nuclear clusters. Compared to that in uniform nuclear matter, the shear viscosity with clusters is smaller at higher temperatures but larger at lower temperatures. The temperature for the minimum specific shear viscosity is largely affected by the energy dependence of NN cross sections, and this is different from the behavior by using a constant NN cross section as seen in Ref.~\cite{Hua:2023amo}. On the other hand, the bulk viscosity increases dramatically with nuclear clusters, and such effect is insensitive to the energy dependence of NN cross sections. Increasing the isospin asymmetry of the system generally increases both viscosities, while the behavior of the viscosities may be qualitatively modified once the isospin asymmetry is large enough to affect significantly the spinodal region. The Coulomb potential reduces the clustering effect but does not change qualitatively the behaviors of the viscosities.

\begin{acknowledgments}
This work is supported by the Strategic Priority Research Program of the Chinese Academy of Sciences under Grant No. XDB34030000, the National Natural Science Foundation of China under Grant Nos. 12375125, 11922514, and 11475243, and the Fundamental Research Funds for the Central Universities.
\end{acknowledgments}

\bibliography{box-viscosity_iso}
\end{document}